\documentclass[vecphys]{svmult}

\usepackage{makeidx}         % allows index generation
\usepackage{graphicx}        % standard LaTeX graphics tool
\usepackage{multicol}        % used for the two-column index
\usepackage[bottom]{footmisc}% places footnotes at page bottom
\makeindex             % used for the subject index
\def\be{\begin{equation}}
\def\ee{\end{equation}}
\def\bea{\begin{eqnarray}}
\def\eea{\end{eqnarray}}
\def\R{{\cal R}}
\def\S{{\cal S}}
\def\bkone{\mathbf k_1}
\def\bktwo{\mathbf k_2}
\def\bkthree{\mathbf k_3}
\def\bkfour{\mathbf k_4}
\def\fNL{{f_{NL}}}
\def\Mp{m_p}

\begin{document}

\title*{Multiple field inflation}
\author{David Wands}
\institute{Institute of Cosmology and Gravitation, Mercantile House,
University of Portsmouth, Portsmouth PO1 2EG, United Kingdom
\texttt{david.wands''AT''port.ac.uk}}
\maketitle

\abstract{Inflation offers a simple model for very early evolution
of our Universe and the origin of primordial perturbations on large
scales. Over the last 25 years we have become familiar with the
predictions of single-field models, but inflation with more than one
light scalar field can alter preconceptions about the inflationary
dynamics and our predictions for the primordial perturbations. I
will discuss how future observational data could distinguish between
inflation driven by one field, or many fields.  As an example, I
briefly review the curvaton as an alternative to the inflaton
scenario for the origin of structure.\footnote{Based on a talk given
at the 22nd IAP Colloquium, ``Inflation +25'', Paris, June 2006}

\section{Introduction}
\label{sec:1}
% Always give a unique label
% and use \ref{<label>} for cross-references
% and \cite{<label>} for bibliographic references
% use \sectionmark{}
% to alter or adjust the section heading in the running head

Inflation provides an attractively simple model for the early
evolution of our Universe, which can produce a large, spatially flat
and largely homogeneous observable universe. It also provides a
source for small primordial perturbations which are the origin of
the large-scale structure in our Universe today. The vacuum
fluctuations of any light scalar field present during inflation can
be swept up by the inflationary expansion to scales much larger than
the Hubble scale.

Inflation is most commonly discussed in terms of a potential energy
which is a function of a single, slowly rolling, scalar field.
Single field, slow-roll inflation produces an almost Gaussian
distribution of adiabatic density perturbations on super-Hubble
scales with an almost scale-invariant spectrum. But supersymmetric
field theories can contain many scalar fields that could play a role
during inflation and string theory, and other higher-dimensional
theories, yield four-dimensional effective actions with many moduli
fields describing the higher dimensional degrees of freedom. One
should be aware of the different possibilities that open up in
particle physics models containing more than one light scalar field
during inflation.

The presence of multiple fields during inflation can lead to quite
different inflationary dynamics, that might appear unnatural in a
single field model, and to spectra of primordial perturbations that
would actually be impossible in single field models. The presence of multiple
light fields during inflation leads to the generation of
non-adiabatic field perturbations during inflation. This can alter
the evolution of the overall curvature perturbation, for instance
leading to detectable non-Gaussianity, and may leave residual
isocurvature fluctuations in the primordial density perturbation on
large scales after inflation, which can be correlated with the
curvature perturbation. Such alternative models are interesting not
only as theoretical possibilities, but because they could be
distinguished by increasingly precise observations in the near
future.

In this short review I will discuss some of the distinctive
observational predictions of inflation in presence of more than one
scalar field. For a more comprehensive review of inflationary
dynamics and reheating with multiple fields see Ref.~\cite{RMP}.

\section{Homogeneous scalar field dynamics}
\label{sec:2}

The time-evolution of a single, spatially homogeneous scalar field
is governed by the Klein-Gordon equation
 \begin{equation}
 \label{singleKG}
\ddot\phi +3H\dot\phi = - \frac{dV}{d\phi} \,,
 \end{equation}
where the Hubble expansion rate is given by the Friedmann constraint
 \begin{equation}
3H^2 = 8\pi G \left[ \frac12 \dot\phi^2 + V(\phi) \right] \,.
 \end{equation}

Multiple scalar fields obey the Klein-Gordon equation
 \begin{equation}
 \label{KG}
\ddot\varphi_I +3H\dot\varphi_I = - \frac{\partial}{\partial\varphi_I}
 \left( \sum_J U_J \right)
 \,,
 \end{equation}
where one must allow for the possibility that the potential energy
is given by a sum over many terms
\begin{equation}
 V = \sum_J U_J \,.
\end{equation}
The wider range of interaction potentials possible in multiple field
models leads to possibilities such as hybrid inflation.

In hybrid inflation models \cite{Linde,Copeland} the {\em inflaton}
field, $\varphi_1$, can roll towards the non-zero minimum of its
potential, $U_1=V_0+m_1^2\varphi_1^2/2$, which would lead to eternal
inflation into the future in a single field model.  But in a hybrid
model there is a second {\em waterfall} scalar field trapped during
inflation in a local minimum, $\varphi_2=0$, with a potential, e.g.,
$U_2=(g^2\varphi_1^2-m_2^2)\varphi_2^2/2$ which becomes unstable below a
critical value of the $\varphi_1$ field, triggering an instability of the
vacuum energy driving inflation, rapidly bringing inflation to an end.

Another more subtle change enters through the Friedmann constraint:
\begin{equation}
 \label{Friedmann}
3H^2 = 8\pi G \left( V + \sum_I \frac12 \dot\varphi_I^2 \right) \,.
 \end{equation}
Even in the absence of explicit interactions in the scalar field
Lagrangian, the fields will still be coupled gravitationally. In
particular the Hubble expansion rate that enters the Klein-Gordon
equation (\ref{KG}) is due to the sum over all fields in
Eq.~(\ref{Friedmann}) and this can also alter the field dynamics even
if the potential for each individual field is left unchanged. The
additional Hubble damping present due to multiple fields can be used
to drive slow-roll inflation in assisted inflation models
\cite{Liddle} where the individual potentials would be too steep
to drive inflation on their own.

The original assisted inflation model \cite{Liddle} considered $n$
scalar fields with steep exponential potentials
\begin{equation}
V = \sum_I U_{I0} \exp \left( -\lambda_I\varphi_I / \Mp \right)
\end{equation}
where I have used the reduced Planck mass $\Mp^2=(8\pi G)^{-1}$. Each
scalar field potential is too steep to drive inflation on its own
if $\lambda_I^2>2$, but the additional damping effect due to the presence
of the other scalar fields leads to a particular power-law inflation
solution, $a\propto t^p$ where $p=2/\lambda^2$ where the
combined fields have an effective potential $V\propto
\exp(-\lambda\sigma/\Mp)$ with
\begin{equation}
\frac{1}{\lambda^2} = \sum_I \frac{1}{\lambda_I^2} \,.
\end{equation}
Thus $\lambda\to0$ for many fields as $n\to\infty$ and we can have
slow-roll inflation even when each $\lambda^2_I>2$.

Even though the background dynamics can be reduced to an equivalent
single field with a specified potential \cite{Malik},
there is an important qualitative difference between the
inflationary dynamics in multiple field inflation with respect to
the single field case. The Hubble damping during inflation drives a
single scalar field to a unique attractor solution during slow-roll
inflation where the Hubble rate, field time-derivative and all local
variables are a function of the local field value: $H(\phi)$,
$\dot\phi(\phi)$, etc. This means that the evolution rapidly becomes
independent of the initial conditions.

In multiple field models we may have a family of trajectories in
phase space where, for example, the Hubble rate at a particular
value of $\varphi_1$ is also dependent upon the value of $\varphi_2$. In
this case the inflationary dynamics, and hence observational
predictions, may be dependent upon the trajectory in phase space and
thus the initial field values. It is this that allows non-adiabatic
perturbations to survive on super-Hubble scales in multiple field
inflation.

It is important to distinguish here between models, such as most
hybrid models, with multiple fields but only one light direction in
field space with small effective mass $\partial^2V/\partial\varphi^2\ll
H^2$ during inflation, and models with many light fields, such as
assisted inflation models. Only models with multiple {\em light}
fields can have multiple slow-roll trajectories.

\subsection{Inflaton field direction during inflation}

It is convenient to identify the {\em inflaton} field direction as the
direction in field space corresponding to the evolution of the
background (spatially homogeneous) fields during inflation
\cite{Gordon00} (see also \cite{Groot01,Rigopoulos}). Thus for $n$
scalar fields $\varphi_I$, where $I$ runs from $1$ to $n$, we have
\begin{equation}
\label{defsigma} \sigma = \int \sum_I \hat\sigma_I \dot\varphi_I
{\rm d}t \,,
\end{equation}
where the inflaton direction is defined by
\begin{equation}
\label{defhatsigma} \hat\sigma_I \equiv
\frac{\dot\varphi_I}{\sqrt{\sum_J \dot\varphi_J^2}}\,.
\end{equation}

The $n$ evolution equations for the homogeneous scalar fields
(\ref{KG}) can then be written as the evolution for a single
inflaton field (\ref{singleKG})
\begin{equation}
\ddot\sigma + 3H\dot\sigma + V_\sigma = 0 \,,
\end{equation}
where the potential gradient in the direction of the inflaton is
\begin{equation}
V_\sigma \equiv \frac{\partial V}{\partial \sigma} = \sum_I
\hat\sigma_I \frac{\partial V}{\partial\varphi_I} \,.
\end{equation}
The total energy density and pressure are then given by the usual single
field results for the inflaton.% (\ref{rhoeq}).

\subsection{An example: Nflation}

A topical example of multiple field inflation is Nflation.
Dimopoulos et al \cite{Nflation} proposed this model based on the
very large number of axion fields predicted in low energy effective
theories derived from string theory. Near the minimum of the
effective potential the fields have a potential energy
\begin{equation}
 \label{VN}
V = \frac12 \sum_I m_I^2 \varphi_I^2 \,.
\end{equation}
This form of potential with a large number of massive fields was
also previously studied by Kanti and Olive \cite{KO} and Kaloper and
Liddle \cite{KL}.

With a single scalar field the quadratic potential yields the
familiar chaotic inflation model with a massive field
$V=m^2\phi^2/2$. But to obtain inflation with a single massive field
the initial value of the scalar field must be several times the
Planck mass and there is a worry that we have no control over
corrections to the potential at super-Planckian values in the
effective field theory \cite{LythRiotto}. But with
many scalar fields the collective dynamics can yield inflation even
for sub-Planckian values if there are a sufficiently large number of
fields. Kim and Liddle \cite{Kim} have found that for random initial
conditions, $-\Mp<\varphi_I(0)<\Mp$, the total number of e-folds is
given by $n/12$, where $n$ is the total number of fields. Thus we
require $n>600$ for sufficient inflation if none of the fields is
allowed to exceed the Planck scale. This may seem to be a large
number, but Dimopoulos et al \cite{Nflation} cite string theory models
with of order $10^5$ axion fields.

As remarked earlier, in the presence of more than one light field,
the trajectory in field space at late times, and hence the
observable predictions, may be dependent upon the initial conditions
for the different fields. But Kim and Liddle \cite{Kim} found
evidence for what they called a ``thermodynamic'' regime where the
predicted spectral index, $n_\R$, for the primordial curvature
perturbations that arise from quantum fluctuations of the scalar
fields, became independent of the precise initial conditions for a
sufficiently large number of fields.
In fact inflation with an arbitrary number of massive fields always
yields a robust prediction for the tensor-to-scalar ratio $r$ in
terms of the number of e-foldings, $N$, from the end of inflation
\cite{LythRiotto}
\begin{equation}
 r = 8 / N \,,
\end{equation}
completely independently of the initial conditions. Thus Nflation
seems to be an example of a multiple field model of inflation which
makes observable predictions which need not depend upon the specific
trajectory in field space.

In the limit where the masses become degenerate, $m_I^2\to m$, the
Nflation dynamics becomes particularly simple. The fields evolve
radially towards the origin and the potential (\ref{VN}) reduces to
that for a single field
\begin{equation}
 V \to \frac12 m^2 \sigma^2 \,,
 \end{equation}
where $\sigma$ is the inflaton field (\ref{defsigma}). Thus in this
limit Nflation reproduces the single field prediction for the
tensor-scalar ratio $r=0.16$ and the spectral index $n_\R=0.96$. However
the presence of $n$ light fields during Nflation also leads to $n-1$
isocurvature modes during inflation and these have an exactly
scale-invariant spectrum (up to first order in the slow-roll
parameters) in the limit of degenerate masses \cite{Byrnes05}. In the
following sections I will describe some of the distinctive predictions
that can arise due to the existence of such non-adiabatic
perturbations during inflation.

\section{Primordial perturbations from inflation}

I have so far only presented equations for the dynamics of homogeneous
scalar fields driving inflation. But to test theoretical predictions
against cosmological observations we need to consider inhomogeneous
perturbations.
It is the primordial perturbations produced during inflation that
offer the possibility of determining the physical processes that
drove the dynamical evolution of the very early universe. In the
standard Hot Big Bang model there seems to be no way to explain the
existence of primordial perturbations, during the radiation
dominated era, on scales much larger than the causal horizon, or
equivalently the Hubble scale. But inflation takes perturbations on
small, sub-Hubble, scales and can stretch them up to arbitrarily
large scales.

\subsection{Scalar field perturbations, without interactions}

Consider an inhomogeneous perturbation,
$\varphi_I\to\varphi_I(t)+\delta\varphi_I(t,{\bf x})$, of the Klein-Gordon
equation (\ref{KG}) for a non-interacting scalar field in an
unperturbed FRW universe. (I will include perturbations of the
metric and other fields later, but for simplicity I will neglect
this complication for the moment.)
 \begin{equation}
\label{simpleKG}
\ddot{\delta\varphi}_I +3H\dot{\delta\varphi}_I + \left( m_I^2 - \nabla^2
\right) \delta\varphi_I = 0 \,,
 \end{equation}
where the effective mass-squared of the field is
$m_I^2=\partial^2V/\partial\varphi_I^2$, and $\nabla^2$ is the spatial
Laplacian. Decomposing an arbitrary field perturbation into
eigenmodes of the spatial Laplacian (Fourier modes in flat space)
$\nabla^2\delta\varphi_I=-(k^2/a^2)\delta\varphi_I$, where $k$ is the
comoving wavenumber and $a$ the FRW scale factor, we find that
small-scale fluctuations in scalar fields on sub-Hubble scales (with
comoving wavenumber $k>aH$) undergo under-damped oscillations, and
on sufficiently small scales are essentially freely oscillating.
Normalising the initial amplitude of these small-scale fluctuations
to the zero-point fluctuations of a free field in flat spacetime we
have \cite{LLKCreview}
\begin{equation}
 \label{deltaphi}
\delta\varphi_I \simeq \frac{e^{-ikt/a}}{a\sqrt{2k}} \,.
\end{equation}
During an accelerated expansion $\dot{a}=aH$ increases and modes
that start on sub-Hubble scales ($k>aH$) are stretched up to
super-Hubble scales ($k<aH$). Perturbations in light fields (with
effective mass-squared $m^2<9H^2/4$) become over-damped (or
``frozen-in'') and Eq.~(\ref{deltaphi}) evaluated when $k\simeq aH$
gives the power spectrum for scalar field fluctuations at
``Hubble-exit''
\begin{equation}
\label{H2pi} {\cal P}_{\delta\varphi_I} \equiv \frac{4\pi k^3}{(2\pi)^3}
\left| \delta\varphi_I^2 \right| \simeq \left( \frac{H}{2\pi} \right)^2
\,.
\end{equation}
Heavy fields with $m^2>9H^2/4$ remain under-damped and have
essentially no perturbations on super-Hubble scales. But light
fields become over-damped and can be treated as essentially
classical perturbations with a Gaussian distribution on super-Hubble
scales.

Thus inflation generates approximately scale-invariant perturbation
spectra on super-Hubble scales in any light field (for $m^2\ll H^2$
and $|\dot{H}|\ll H^2$).

\subsection{Scalar field and metric perturbations, with interactions}

The simplified discussion in the preceding subsection gives a good
approximation to the scalar field perturbations generated around the
time of Hubble-exit during slow-roll inflation, where field
interactions and metric backreaction are small. However to accurately
track the evolution of perturbations through to the end of inflation
and into the radiation dominated era we need to include interactions
between fields and, even in the absence of explicit interactions,
we need to include gravitational backreaction.

%\subsection{Metric perturbations}

For an inhomogeneous matter distribution the Einstein equations
imply that we must also consider inhomogeneous metric perturbations
about the spatially flat FRW metric. The perturbed FRW spacetime is
described by the line element~\cite{Mukhanov90,RMP}
\begin{eqnarray}
\hspace*{-0.2em} ds^2 &=& - (1+2A) dt^2 + 2a\partial_iB dx^i dt
\nonumber\\
\hspace*{-0.2em}&& +a^2\left[ (1-2\psi)\delta_{ij} + 2\partial_{ij}E
 + h_{ij} \right] dx^i dx^j\,,
\end{eqnarray}
where $\partial_i$ denotes the spatial partial derivative
$\partial/\partial x^i$. We will use lower case latin indices to run
over the 3 spatial coordinates.
%Our metric perturbations follow the
%notation of Ref.~\cite{Mukhanov90}, apart from our use of $A$ rather
%than $\phi$ as the perturbation in the lapse function.

The metric perturbations have been split into scalar and tensor parts
according to their transformation properties on the spatial
hypersurfaces. The field equations for the scalar and tensor parts
then decouple to linear order. Vector metric perturbations are
automatically zero at first order if the matter content during
inflation is described solely by scalar fields.

%\subsubsection{Tensor modes}

The tensor perturbations, $h_{ij}$, are transverse ($\partial^i
h_{ij}=0$) and trace-free ($\delta^{ij}h_{ij}=0$). They are
automatically independent of coordinate gauge transformations.
These describe gravitational waves as they are the free part of the
gravitational field and evolve independently of linear matter
perturbations.

We can decompose arbitrary tensor perturbations into eigenmodes of
the spatial Laplacian, $\nabla^2e_{ij}=-(k^2/a^2)e_{ij}$, with comoving
wavenumber $k$, and scalar amplitude $h(t)$:
\begin{equation}
\label{eq:defh} h_{ij} = h(t) e_{ij}^{(+,\times)}(x)\,,
\end{equation}
with two possible polarisation states, $+$ and $\times$.
The Einstein equations yield a wave equation for the amplitude of the
tensor metric perturbations
\begin{equation}
 \label{teneq}
\ddot{h} + 3H\dot{h} + \frac{k^2}{a^2} h = 0 \,,
\end{equation}
This is the same as the wave equation (\ref{simpleKG}) for a massless
scalar field in an unperturbed FRW metric.

%\subsubsection{Scalar perturbations}

The four scalar metric perturbations $A$, $\partial_iB$,
$\psi\delta_{ij}$ and $\partial_{ij}E$ are constructed from
3-scalars, their derivatives, and the background spatial metric. The
intrinsic Ricci scalar curvature of constant time hypersurfaces is
given by
\begin{equation}
^{(3)}R = \frac{4}{a^2} \nabla^2 \psi \,.
\end{equation}
Hence we refer to $\psi$ as the curvature perturbation.

% Under a scalar coordinate/gauge transformation
%%
%\begin{eqnarray}
%t &\to& t+ \delta t\,, \\
%x^i &\to& x^i + \delta^{ij} \partial_j \delta x\,,
%\end{eqnarray}
%%
%$\delta t$ determines the time slicing and $\delta x$ the spatial
%threading. The scalar metric perturbations then transform as
%%
%\begin{eqnarray}
%A &\to& A -\dot{\delta t} \,, \\
%B &\to& B + a^{-1} \delta t - a\dot{\delta x}  \,,\\
%E &\to& E - \delta x \,,\\
%\psi &\to& \psi + H\delta t \,.
%\end{eqnarray}
%%
%Although $B$ and $E$ separately are spatially gauge-dependent, the
%combination $\dot{E}-B/a$ is independent of spatial gauge and
%describes the scalar potential for the anisotropic shear of
%worldlines orthogonal to constant time hypersurfaces
%\cite{Kodama84}.

First-order scalar field perturbations in a first-order perturbed
FRW universe obey the wave equation~\cite{RMP}
\begin{eqnarray}
\label{eq:pertKG}
\label{eq:scalareom}
\hspace*{-2.0em}&&
\ddot{\delta\varphi}_I + 3H\dot{\delta\varphi}_I
 + \frac{k^2}{a^2} \delta\varphi_I + \sum_J V_{IJ}
\delta\varphi_J
  \nonumber\\
  \hspace*{-2.0em}&&~~{}
=
-2V_{I}A + \dot\varphi_I \left[ \dot{A} + 3\dot{\psi} +
\frac{k^2}{a^2} (a^2\dot{E}-aB) \right]. \label{eq:perturbation}
\end{eqnarray}
where the mass-matrix $V_{IJ}\equiv\partial^2
V/\partial\varphi_I\partial\varphi_J$.
The Einstein equations relate the scalar metric perturbations to matter
perturbations via the energy and momentum constraints \cite{Mukhanov90}
\begin{eqnarray}
\label{eq:densitycon}
 3H\left(\dot\psi+HA\right) +
\frac{k^2}{a^2}\left[\psi+H(a^2\dot{E}-aB)\right]
 &=&-4\pi G \delta\rho \,, \\
 \dot\psi + HA &=& -4\pi G \delta q \,, \label{eq:mtmcon}
\end{eqnarray}
where the energy and pressure perturbations and momentum for $n$ scalar fields
are given by~\cite{RMP}
\begin{eqnarray}
\delta\rho &=& \sum_I\left[
 \dot\varphi_I \left( \dot{\delta\varphi}_I -\dot\varphi_I A \right)
 + V_{I}\delta\varphi_I \right] \,,
\label{eq:densityphi} \\
\delta P &=& \sum_I\left[
 \dot\varphi_I \left( \dot{\delta\varphi}_I -\dot\varphi_I A \right)
 - V_{I}\delta\varphi_I \right] \,,
\label{eq:pressurephi} \\
\delta q_{,i} &=& - \sum_I \dot{\varphi}_I
\delta\varphi_{I,i} \,,
\label{eq:mtmphi}
\end{eqnarray}
where $V_I \equiv \partial V/\partial \varphi_I$.

%\subsection{Gauge-invariant variables}

We can construct a variety of gauge-invariant combinations of the
scalar metric perturbations. The longitudinal gauge corresponds to a
specific gauge-transformation to a (zero-shear) frame such that
$E=B=0$, leaving the gauge-invariant variables
\begin{eqnarray}
 \label{defPhi}
\Phi &\equiv& A - \frac{{\rm d}}
{{\rm d}t} \left[ a^2(\dot{E}-B/a)\right] \,,\\
 \label{defPsi}
\Psi &\equiv& \psi + a^2 H (\dot{E}-B/a) \,.
\end{eqnarray}
%

%Matter perturbations are also gauge-dependent. Scalar field, density
%and pressure perturbations all obey the simple transformation rule
%%
%\begin{equation}
%\delta\rho \to \delta\rho - \dot\rho \delta t \,.
%\end{equation}
%%
%The adiabatic pressure perturbation is defined to be
%%
%\begin{equation}
%\delta P_{\rm ad} \equiv \frac{\dot P}{\dot\rho} \delta\rho \,,
%\end{equation}
%%
%and hence the non-adiabatic part of the actual pressure
%perturbation, or entropy perturbation, is a gauge-invariant
%perturbation
%%
%\begin{equation}
% \label{defPnad}
%\delta P_{\rm nad} = \delta P -\frac{\dot P}{\dot\rho} \delta\rho\,.
%\end{equation}
%%
%The scalar part of the 3-momentum is given by $\partial_i\delta q$
%and this momentum potential transforms as
%%
%\begin{equation}
%\delta q \to \delta q + (\rho+P)\delta t \,.
%\end{equation}
%%
%Thus we can obtain the gauge-invariant comoving density
%perturbation~\cite{Bardeen80}
%%
%\begin{equation}
% \label{defrhom}
%\delta\rho_m = \delta\rho - 3H \delta q \,.
%\end{equation}
%

Another variable commonly used to describe scalar
perturbations during inflation is the field perturbation in the
spatially flat gauge (where $\psi=0$). This has the gauge-invariant
definition \cite{Mukhanov85,Sasaki86}:
\begin{equation}
 \label{eq:defdphipsi}
 \delta\varphi_{I\psi} \equiv \delta\varphi_I + \frac{\dot\varphi}{H} \psi \,.
\end{equation}
%
%In single field inflation this is simply a rescaling of the comoving
%curvature perturbation ${\cal R}$ in (\ref{eq:defR}). We see that
%what appears as a field perturbation in one gauge is a metric
%perturbation in another gauge and vice versa.
%
It is possible to use the Einstein equations to eliminate the metric
perturbations from the perturbed Klein-Gordon equation
(\ref{eq:pertKG}), and write a wave equation
solely in terms of the field perturbations in the spatially flat gauge
\cite{Taruya:1997iv}
\begin{equation}
   \ddot{\delta\varphi}_{I\psi}
 + 3H\dot{\delta\varphi}_{I\psi}
 + \frac{k^2}{a^2}\delta\varphi_{I\psi} +
  \sum_J
   \left[ V_{IJ} -\frac{8\pi G}{a^3}
 \frac{d}{dt} \left( \frac{a^3}{H} \dot{\varphi}_I\dot{\varphi}_J
 \right) \right]
   \delta\varphi_{J\psi}=0
 \,.
 \end{equation}

Only at lowest order in the slow-roll expansion can the interaction
terms be neglected and we recover the simplified wave equation
(\ref{simpleKG}) for a massless field in an unperturbed FRW universe.

\subsection{Adiabatic and entropy perturbations}

There are two more gauge-invariant scalars which are commonly used to
describe the overall curvature perturbation.
The curvature perturbation on uniform-density hypersurfaces is given
by
\begin{equation}
 \label{defzeta}
- \zeta \equiv \psi + \frac{H}{\dot\rho}\delta\rho \,,
\end{equation}
first introduced by Bardeen, Steinhardt and Turner~\cite{Bardeen83}
(see also Refs.~\cite{Bardeen88,Martin98,Wands00}).
The comoving curvature perturbation (strictly speaking the curvature
perturbation on hypersurfaces orthogonal to comoving worldlines)
\begin{equation}
 \label{eq:defR}
{\cal R} \equiv \psi - \frac{H}{\rho+P} \delta q \,,
\end{equation}
where the scalar part of the 3-momentum is given by
$\partial_i\delta q$.
${\cal R}$ has been used by Lukash \cite{Lukash80}, Lyth
\cite{Lyth85} and many others.
%, including Mukhanov, Feldman and
%Brandenberger in their review \cite{Mukhanov90}.
%
%(Note that in their review the comoving curvature perturbation is
%denoted by ``$\zeta$'' in Ref.~\cite{Mukhanov90} and defined in
%terms of the metric perturbations in the longitudinal gauge, but it
%is equivalent to our definition of ${\cal R}$ in a spatially flat
%background with vanishing anisotropic stress.)
%
For single field inflation we have $\delta q=-\dot\phi\delta\phi$
and hence
\begin{equation}
{\cal R} = \psi + \frac{H}{\dot\phi}\delta\phi \,.
\end{equation}

The difference between the two curvature perturbations $-\zeta$ and
${\cal R}$,
\begin{equation}
\label{eq:zetaR} -\zeta - {\cal R} = \frac{H}{\dot\rho} \delta\rho_m
\,,
\end{equation}
is proportional to the comoving density
perturbation~\cite{Bardeen80}
\begin{equation}
 \label{defrhom}
\delta\rho_m \equiv \delta\rho - 3H \delta q \,.
\end{equation}
%
%In slow-roll single-field inflation we have $\delta\rho/\dot\rho
%\simeq \delta\phi/\dot\phi$ and hence $\delta\rho_m\simeq0$ and
%these two commonly used curvature perturbations, ${\cal R}$ and
%$-\zeta$, coincide.
%More generally
The energy and momentum constraints (\ref{eq:densitycon}) and
(\ref{eq:mtmcon}) can be combined to give a generalisation of the Poisson
equation
\begin{equation}
 \label{eq:rhomcon}
%\frac{k^2}{a^2} \Psi = - 4\pi G \delta\rho_m \,,
 \frac{\delta\rho_m}{\rho} = - \frac23 \left( \frac{k}{aH} \right)^2
 \Psi \,,
\end{equation}
relating the longitudinal gauge metric perturbation (\ref{defPsi}) to
the comoving density perturbation (\ref{defrhom}). Thus the two
curvature perturbations, ${\cal R}$ and $-\zeta$, coincide on large
scales ($k/aH\ll1$) so long as the longitudinal gauge metric
perturbation, $\Psi$, remains finite - which is generally true during
slow-roll inflation.

The energy conservation equation can be written in terms
of the curvature perturbation on uniform-density hypersurfaces,
defined in (\ref{defzeta}), to obtain the first-order evolution
equation~\cite{Wands00,RMP}
\begin{equation}
\label{eq:dotzeta}
\dot\zeta = -H \frac{\delta P_{\rm nad}}{\rho+P} - {\Sigma} \,,
\end{equation}
where $\delta P_{\rm nad}$ is the
non-adiabatic pressure perturbation,
% defined in (\ref{defPnad}),
%
\begin{equation}
 \label{defPnad}
\delta P_{\rm nad} = \delta P -\frac{\dot P}{\dot\rho}
\delta\rho \,,
\end{equation}
and $\Sigma$ is
the scalar shear along comoving worldlines~\cite{Lyth03p1},
which can be given relative to the Hubble rate as
\begin{eqnarray}
\hspace*{-0.5em} \frac{\Sigma}{H} &\equiv& - \frac{k^2}{3H} \left\{
\dot{E}-(B/a) + \frac{\delta
    q}{a^2(\rho+P)} \right\} \nonumber \\
\hspace*{-0.5em} &=& - \frac{k^2}{3a^2H^2} \zeta
 - \frac{k^2 \Psi}{3a^2H^2} \left[ 1 - \frac{2\rho}{9(\rho+P)}
     \frac{k^2}{a^2H^2} \right] \,.
\end{eqnarray}
Thus $\zeta$ and $\R$ are constant (and equal upto a sign difference)
for adiabatic perturbations on super-Hubble scales ($k/aH\ll1$), so
long as $\Psi$ remains finite, in which case the shear of comoving
worldlines can be neglected.

More generally we can define adiabatic perturbations to be
perturbations which lie along the background trajectory in the phase
space of spatially homogeneous fields \cite{Wands00,Gordon00}. That is, we
generalise Eq.~(\ref{defPnad}) so that for adiabatic linear
perturbations of any two variables $x$ and $y$ we require
\begin{equation}
 \frac{\delta x}{\dot{x}} = \frac{\delta y}{\dot{y}} \,: \quad {\rm
 adiabatic}
 \end{equation}
Thus adiabatic perturbations in a multiple field inflation can be
characterised by a unique shift along the background trajectory
$\delta N = -H \delta x/\dot{x}= -H \delta y/\dot{y}$.
For example, for adiabatic perturbations of the primordial plasma we
require that the baryon-photon ratio, $n_B/n_\gamma$, remains
unperturbed, and hence
\begin{equation}
 \frac{\delta(n_B/n_\gamma)}{n_B/n_\gamma}
  = -3H \left( \frac{\delta n_B}{\dot{n}_B} - \frac{\delta
  n_\gamma}{\dot{n}_\gamma} \right)
   = 0 \,,
\end{equation}
where we have used $\dot{n}_x=-3Hn_x$ for baryon number density, and
photon number density below about 1~MeV.

For a single scalar field the non-adiabatic pressure (\ref{defPnad})
can be related to the comoving density perturbation (\ref{defrhom})
\cite{Gordon00}
\begin{equation}
\delta P_{\rm nad} = - \frac{2V_{,\varphi}}{3H\dot\varphi}
\delta\rho_m \,.
\end{equation}
{}From the Einstein constraint (\ref{eq:rhomcon}) this will
vanish on large scales ($k/aH\to0$) if $\Psi$ remains finite, and
hence single scalar field perturbations become adiabatic in this
large-scale limit.
In particular, we have from Eq.~(\ref{eq:dotzeta}) that $\zeta$ becomes
constant for adiabatic perturbations in this large scale limit, and
hence the curvature perturbation can be calculated shortly after
Hubble-exit in single field inflation and equated directly with the
primordial curvature perturbation, independently of the details of
reheating, etc, at the end of inflation.
But in the presence of more than one light field the vacuum
fluctuations stretched to super-Hubble scales will inevitably
include non-adiabatic perturbations due to the presence of multiple
trajectories in the phase space.

We define entropy perturbations to be fluctuations orthogonal to the
background trajectory
\begin{equation}
S_{xy} \propto \frac{\delta x}{\dot{x}} - \frac{\delta y}{\dot{y}}
\,: \quad {\rm
 entropy}
\end{equation}
For example in the primordial era we can have entropy perturbations
in the primordial plasma
\begin{equation}
 \label{defSB}
 S_B = \frac{\delta(n_B/n_\gamma)}{n_B/n_\gamma}
  = - 3H \left( \frac{\delta n_B}{\dot{n}_B} - \frac{\delta
  n_\gamma}{\dot{n}_\gamma} \right) \,.
\end{equation}
which are also referred to as baryon isocurvature perturbations.

In the radiation-dominated era we can define a gauge-invariant
primordial curvature perturbations associated with each of the
component fluids \cite{Wands00,MW} in analogy with the total curvature
perturbation (\ref{defzeta})
\begin{equation}
 \label{defzetaI}
\zeta_I \equiv -\psi - H \frac{\delta\rho_I}{\dot\rho_I} \,,
\end{equation}
and these will be constant in the large-scale limit for
non-interacting fluids with barotropic equation of state $P_i(\rho_i)$
(and hence vanishing non-adiabatic pressure perturbations for each
fluid) \cite{Wands00}. We can identify isocurvature perturbations, such as the
baryon isocurvature perturbation (\ref{defSB}), with the difference
between each $\zeta_I$ and, by convention, $\zeta_\gamma$ for the photons
\begin{equation}
 \label{defSI}
 S_I \equiv 3 \left( \zeta_I - \zeta_\gamma \right) \,,
\end{equation}
and the total curvature perturbation (\ref{defzeta}) is given by the
weighted sum
\begin{equation}
 \label{zetasum}
\zeta = \sum_I \frac{\dot\rho_I}{\dot\rho} \zeta_I \,.
\end{equation}

Scale-invariant spectra of primordial isocurvature perturbations
give rise to distinctive power spectra of CMB temperature
anisotropies and polarisation \cite{Moodley} and are thus tightly
constrained by observations \cite{Juan,Bean,Trotta}, although there
may be hints of isocurvature modes in current CMB data \cite{Jussi}.

Arbitrary field perturbations in multiple field inflation can be
decomposed into adiabatic perturbations along the inflaton
trajectory and $n-1$ entropy perturbations orthogonal to the
inflaton direction (\ref{defhatsigma}) in field space:
\begin{eqnarray}
\label{defdeltasigma}
\delta\sigma &=& \sum_I \hat\sigma_I \delta\varphi_I \,, \\
\label{defdeltas} \delta s_{I} &=& \sum_J \hat{s}_{IJ}
\delta\varphi_J \,,
\end{eqnarray}
where $\sum_I \hat{s}_{JI}\hat\sigma_I=0$.
Without loss of generality I will assume that the entropy fields are
also mutually orthogonal in field space. Note that I have assumed
that the fields have canonical kinetic terms, that is, the field
space metric is flat. See Refs.~\cite{Groot01,DiMarco03,Choi07} for the
generalisation to non-canonical kinetic terms.

The total momentum and pressure perturbation (\ref{eq:mtmphi}) and
(\ref{eq:pressurephi}) for $n$ scalar field perturbations can be
written in the same form as for a single inflaton field
\begin{eqnarray}
\delta q &=& - \dot\sigma \delta\sigma \,, \\
\delta P &=& \dot\sigma(\dot{\delta\sigma}-\dot\sigma A) -
V_\sigma\delta\sigma \,.
\end{eqnarray}
However the density perturbation (\ref{eq:densityphi}) is given by
\begin{equation}
\delta\rho = \dot\sigma(\dot{\delta\sigma}-\dot\sigma A) +
V_\sigma\delta\sigma + 2 \delta_s V \,,
\end{equation}
where the deviation from the single field result
arises due to the non-adiabatic perturbation of the
potential orthogonal to the inflaton trajectory:
\begin{equation}
\delta_s V \equiv \sum_I V_I \delta\varphi_I - V_\sigma \delta\sigma
 \,.
\end{equation}
The non-adiabatic pressure perturbation (\ref{defPnad}) is written
as \cite{Gordon00,RMP}
\begin{equation}
\delta P_{\rm nad} = - \frac{2V_\sigma}{3H\dot\sigma} \delta\rho_m -
2\delta_s V \,,
\end{equation}
where the comoving density perturbation, $\delta\rho_m$, is given by
Eq.~(\ref{defrhom}).
Although the constraint Eq.~(\ref{eq:rhomcon}) requires the comoving
density perturbation to become small on large scales, as in the single
field case, there is now an additional contribution to the
non-adiabatic pressure due to non-adiabatic perturbations of the
potential which need not be small on large scales.
%We note that the
%dynamics of cosmological perturbations in multi-field system was
%investigated by a host of authors, see
%Refs.~\cite{Bartolo01p1,Chiba97,Linde85,Salopek89,Sta94,Sta01,FRV04,Garcia95,Garcia96,HY05,Hwang00,Kadota03,Kadota03v2,Kana00p1,Kawasaki01,Polarski94,Sasaki95,Sasaki98,Langlois99,Langlois00,Mukhanov97,Kofman87,Kofman88,Lahiri}.

It is important to emphasise that the presence of entropy
perturbations during inflation does not mean that the ``primordial''
density perturbation (at the epoch of primordial nucleosynthesis) will
contain isocurvature modes. In particular, if the universe undergoes
conventional reheating at the end of inflation and all particle
species are driven towards thermal equilibrium with their abundances
determined by a single temperature (with no non-zero chemical
potentials) then the primordial perturbations must be adiabatic
\cite{Weinberg04b}.  It is these primordial perturbations that set the
initial conditions for the evolution of the radiation-matter fluid that
determines the anisotropies in the cosmic microwave background and
large-scale structure in our Universe, and thus are directly
constrained by observations.
We will see that while the existence of non-adiabatic perturbations
after inflation requires the existence of non-adiabatic
perturbations during inflation \cite{Weinberg04a}, it is not true
that non-adiabatic modes during inflation necessarily give primordial
isocurvature modes \cite{Weinberg04b}.

%%%%%%%%%%%%%%%%%%%%%%%%%%%%%%%%%%%%%%%%
%\section{Correlated spectra of adiabatic and entropy perturbations}
\section{Perturbations from two-field inflation}
%%%%%%%%%%%%%%%%%%%%%%%%%%%%%%%%%%%%%%%%

In this section I will consider the specific example of the coupled
evolution of two canonical scalar fields, $\phi$ and $\chi$, during
inflation and how this can give rise to correlated curvature and
entropy perturbations on large scales after inflation
\cite{Langlois99}.  I will use the local rotation in field space
defined by Eq.~(\ref{defdeltasigma}) and (\ref{defdeltas}) to describe
the instantaneous adiabatic and entropy field perturbations.

The inflaton field perturbation~(\ref{defdeltasigma}) is
gauge-dependent, but we can choose to work with the inflaton
perturbation in the spatially flat ($\psi=0$) gauge:
\begin{equation}
\delta\sigma_\psi \equiv \delta\sigma + \frac{\dot\sigma}{H}\psi \,.
\end{equation}
On the other hand, the orthogonal entropy perturbation (\ref{defdeltas}) is
automatically gauge-invariant.

The generalisation to two fields of the evolution equation for the
inflaton field perturbation in the spatially flat gauge, obtained
from the perturbed Klein-Gordon equations (\ref{eq:pertKG}), is
\cite{Gordon00}
\begin{eqnarray}
\label{eq:twoscalareom}
\hspace*{-1.3em}
&& \ddot{\delta\sigma}_\psi + 3H\dot{\delta\sigma}_\psi
 + \left[ \frac{k^2}{a^2} + V_{\sigma\sigma}
  - \dot{\theta}^2
  - \frac{8\pi G}{a^3} \frac{{\rm d}}
  {{\rm d}t} \left( \frac{a^3\dot\sigma^2}{H} \right)
 \right] \delta\sigma_\psi \nonumber\\
 \hspace*{-1.3em}
 && =  ~~~ 2\frac{{\rm d}}{{\rm d}t}
(\dot\theta\delta s) - 2\left(
{V_\sigma \over \dot\sigma} + {\dot{H}\over H} \right)
\dot\theta \delta s\,,
\end{eqnarray}
and the entropy perturbation obeys
\begin{equation}
\label{eq:entropyeom}
\ddot{\delta s} + 3H\dot{\delta s} + \left(\frac{k^2}{a^2}
  + V_{ss} + 3\dot{\theta}^2 \right) \delta s =
{\dot\theta\over\dot\sigma} {k^2 \over 2\pi G a^2} \Psi\,,
\end{equation}
where $\tan \theta=\dot{\chi}/\dot{\phi}$ and
\begin{eqnarray}
\hspace*{-1.0em}
V_{\sigma \sigma} &\equiv& (\cos^2 \theta) V_{\phi \phi} +(\sin
2\theta)V_{\phi \chi}+(\sin^2 \theta) V_{\chi \chi},\\
\hspace*{-1.0em}
V_{ss} &\equiv& (\sin^2 \theta) V_{\phi \phi}
-(\sin 2\theta)V_{\phi \chi}+(\cos^2 \theta) V_{\chi \chi}.
\label{Vdd}
\end{eqnarray}

We can identify a purely adiabatic mode where $\delta s=0$ on large
scales. However a non-zero entropy perturbation does appear as a
source term in the perturbed inflaton equation whenever the inflaton
trajectory is curved in field space, i.e., $\dot{\theta} \ne 0$.
We note that $\dot{\theta}$ is given by \cite{Gordon00}
\begin{equation}
\dot\theta = - \frac{V_s}{\dot\sigma} \,,
\end{equation}
where $V_s$ is the potential gradient orthogonal to the inflaton
trajectory in field space.

The entropy perturbation evolves independently of the curvature
perturbation on large-scales. It couples to the curvature perturbation
only through the gradient of the longitudinal gauge metric potential,
$\Psi$. Thus entropy perturbations are also described as
``isocurvature'' perturbations on large scales.
Eq.~(\ref{eq:twoscalareom}) shows that the entropy perturbation
$\delta s$ works as a source term for the adiabatic perturbation.
This is in fact clearly seen if we take the time derivative of the
curvature perturbation \cite{Gordon00}:
\begin{equation}
\label{dotR}
\dot{\cal R}=\frac{H}{\dot{H}}\frac{k^2}{a^2}\Psi
+\frac{2H}{\dot{\sigma}}\dot{\theta}\delta s\,.
\end{equation}
Therefore ${\cal R}$ (or $\zeta$) is not conserved even in the
large-scale limit in the presence of the entropy perturbation $\delta
s$ with a non-straight trajectory in field space ($\dot{\theta} \ne
0$).

Analogous to the single field case we can introduce slow-roll
parameters for light, weakly coupled fields \cite{Wands02}. At
first-order in a slow-roll expansion, the inflaton rolls directly down
the potential slope, that is $V_s\simeq0$. Thus we have only one slope
parameter
\begin{equation}
\epsilon \equiv -\frac{\dot{H}}{H^2} \simeq \frac{1}{16\pi G}
\left( \frac{V_\sigma}{V} \right)^2 \,,
\end{equation}
but three parameters, $\eta_{\sigma\sigma}$, $\eta_{\sigma s}$ and
$\eta_{ss}$,
describing the curvature of the potential, where
\begin{equation}
\eta_{IJ} \equiv \frac{1}{8\pi G} \frac{V_{IJ}}{V} \,.
\end{equation}

The background slow-roll solution is described in terms of the
slow-roll parameters by
\begin{equation}
 \label{twofieldbackground}
\dot\sigma^2 \simeq \frac23 \epsilon V \, , \quad
H^{-1} \dot\theta \simeq -\eta_{\sigma s} \,,
\end{equation}
while the perturbations obey
\begin{eqnarray}
\label{twofieldperturbations} H^{-1} \dot{\delta\sigma}_\psi
&\simeq&
 \left( 2\epsilon -\eta_{\sigma\sigma} \right) \delta\sigma_\psi
 - 2\eta_{\sigma s}\delta s \,,
  \nonumber\\
H^{-1} \dot{\delta s} &\simeq& -\eta_{ss}\delta s \,,
\end{eqnarray}
on large scales, where we neglect spatial gradients.
Although $V_s\simeq 0$ at lowest order in slow-roll, this does not
mean
that the inflaton and entropy perturbations decouple. $\dot\theta$
given by Eq.~(\ref{twofieldbackground}) is in general non-zero at
first-order in slow-roll and large-scale entropy perturbations do affect the
evolution of the adiabatic perturbations when $\eta_{\sigma s}\neq 0$.

While the general solution to the two second-order perturbation
equations (\ref{eq:twoscalareom}) and (\ref{eq:entropyeom}) has four
independent modes, the two first-order slow-roll equations
(\ref{twofieldperturbations}) give the approximate form of the
squeezed state on large scales. This has only two modes which we can
describe in terms of dimensionless curvature and isocurvature
perturbations:
\begin{equation}
\label{eq:defRS}
\R \equiv \frac{H}{\dot\sigma} \delta\sigma_\psi \,, \quad
\S \equiv \frac{H}{\dot\sigma} \delta s \,.
\end{equation}
The normalisation of $\R$ coincides with the standard definition of
the comoving curvature perturbation, Eq.~(\ref{eq:defR}). The
normalisation of the dimensionless entropy during inflation, $\S$, is
chosen here
coincide with Ref.~\cite{Wands02}. It can be related to the
non-adiabatic pressure perturbation (\ref{defPnad}) on large scales
\begin{equation}
\delta P_{\rm nad} \simeq - \epsilon
\eta_{\sigma s} \frac{H^2}{2\pi G} \S \,.
\end{equation}

The slow-roll approximation can provide a useful approximation to the
instantaneous evolution of the fields and their perturbations on large
scales during slow-roll inflation, but is not expected to remain
accurate when integrated over many Hubble times, where inaccuracies
can accumulate. In single-field inflation the constancy of the
comoving curvature perturbation after Hubble exit, which does not rely
on the slow-roll approximation, is crucial in order to make accurate
predictions of the primordial perturbations using the slow-roll
approximation only around Hubble crossing. In a two-field model we
must describe the evolution after Hubble exit in terms of
a general transfer matrix:
\begin{equation}
\label{defTransfer}
\left(
\begin{array}{c}
{\R} \\ {\S}
\end{array}
\right) = \left(
\begin{array}{cc}
1 & {T}_{\R\S} \\ 0 & {T}_{\S\S}
\end{array}
\right) \left(
\begin{array}{c}
\R \\\S
\end{array}
\right)_* \,.
\end{equation}
On large scales the comoving curvature perturbation still remains
constant for the purely adiabatic mode, corresponding to $\S=0$,
and adiabatic perturbations remain adiabatic. These general results
are enough to fix two of the coefficients in the transfer matrix, but
${T}_{\R\S}$ and ${T}_{\S\S}$ remain to be determined
either within a given theoretical model,
or from observations, or ideally by both.
The scale-dependence of the transfer functions depends upon
the inflaton-entropy coupling at Hubble exit during inflation and can
be given in terms of the slow-roll parameters as \cite{Wands02}
\begin{eqnarray}
\label{dTdlnk}
\frac{\partial}{\partial\ln k} {T}_{\R\S} &=& 2\eta_{\sigma s} +
 (2\epsilon - \eta_{\sigma\sigma} + \eta_{ss}) {T}_{\R\S}
\,,\nonumber\\
\frac{\partial}{\partial\ln k} {T}_{\S\S} &=&
 (2\epsilon - \eta_{\sigma\sigma} + \eta_{ss}) {T}_{\S\S} \,.
\end{eqnarray}

\subsection{Initial power spectra}

For weakly-coupled, light fields we can neglect interactions on
wavelengths below the Hubble scale, so that vacuum fluctuations give
rise to a spectrum of uncorrelated field fluctuations on the Hubble
scale ($k=aH$) during inflation given by Eq.~(\ref{H2pi}):
\begin{equation}
{\cal P}_{\delta\phi} \simeq {\cal P}_{\delta\chi}
 \simeq \left( \frac{H}{2\pi} \right)_*^2 \,,
\end{equation}
where we use a $*$ to denote quantities evaluated at Hubble-exit.
If a field has a mass comparable to the Hubble scale or larger then
the vacuum fluctuations on wavelengths greater than the effective
Compton wavelength are suppressed. In addition fluctuations in
strongly interacting fields may develop correlations before Hubble
exit. But during slow-roll inflation the correlation between vacuum
fluctuations in weakly coupled, light fields at Hubble-exit is
suppressed by slow-roll parameters. This remains true under a local
rotation in fields space to another orthogonal basis such as the
instantaneous inflaton and entropy directions (\ref{defdeltasigma})
and (\ref{defdeltas}) in field space.

The curvature and isocurvature power spectra at
Hubble-exit are given by
\begin{equation}
 \label{Rstar}
\left. {\cal P}_{\R} \right|_* \simeq
\left. {\cal P}_{\S} \right|_* \simeq
 \left( \frac{H^2}{2\pi\dot\sigma} \right)_*^2
 \simeq \frac{8}{3} \left( \frac{V}{\epsilon M_{\rm Pl}^4} \right)_*
 \,,
\end{equation}
while the cross-correlation is first-order in slow-roll
\cite{vanTent,Byrnes:2006fr},
\begin{equation}
\left. {\cal C}_{\R\S} \right|_* \simeq -2C\eta_{\sigma s} \left.
{\cal P}_{\R} \right|_* \,,
\end{equation}
where $C=2-\ln2-\gamma\approx0.73$ and $\gamma$ is the Euler number.
The normalisation chosen for the dimensionless entropy perturbation in
Eq.~(\ref{eq:defRS}) ensures that the curvature and isocurvature
fluctuations have the same power at horizon exit \cite{Wands02}.
The spectral tilts at horizon-exit are also the same and are given by
\begin{equation}
\label{nR*}
\Delta n_\R|_* \simeq \Delta n_\S|_* \simeq -6\epsilon + 2\eta_{\sigma\sigma} \,.
\end{equation}
where $\Delta n_X \equiv d \ln{\cal P}_X/d \ln k$.

The tensor perturbations (\ref{teneq}) are decoupled from scalar
metric perturbations at first-order and hence the power spectrum has
the same form as in single field inflation. Thus the power spectrum
of gravitational waves on super-Hubble scales during inflation is
given by
\begin{equation}
 \label{Tstar}
\left. {\cal P}_{\rm T} \right|_*
 \simeq \frac{16H^2}{\pi M_{\rm Pl}^2}
 \simeq \frac{128}{3} \frac{V_*}{M_{\rm Pl}^4} \,,
\end{equation}
and the spectral tilt is
\begin{equation}
 \label{nT}
\Delta n_{\rm T}|_* \simeq -2 \epsilon \,.
\end{equation}

\subsection{Primordial power spectra}

The resulting primordial power spectra on large scales can be obtained
simply by applying the general transfer matrix (\ref{defTransfer}) to
the initial scalar perturbations. The scalar power spectra probed by
astronomical observations are thus given by \cite{Wands02}
\begin{eqnarray}
\label{PR}
{\cal P}_\R &=& (1+T_{\R\S}^2) {\cal P}_\R|_* \\
{\cal P}_\S &=& T_{\S\S}^2 {\cal P}_\R|_* \\
\label{CRS}
{\cal C}_{\R\S} &=& T_{\R\S} T_{\S\S} {\cal P}_\R|_* \,.
\end{eqnarray}
The cross-correlation can be given in terms of a dimensionless
correlation angle:
\begin{equation}
\label{defDelta}
\cos\Theta \equiv \frac{{\cal C}_{\R\S}}{\sqrt{{\cal P}_\R{\cal
      P}_\S}} = \frac{T_{\R\S}}{\sqrt{1+T_{\R\S}^2}} \,.
\end{equation}

We see that if we can determine the dimensionless correlation angle,
$\Theta$, from observations, then this determines the off-diagonal
term in the transfer matrix
\begin{equation}
\label{defTRS}
T_{\R\S} = \cot\Theta \,,
\end{equation}
and we can in effect measure the contribution of the entropy
perturbation during two-field inflation to the resultant curvature
primordial perturbation. In particular this allows us in principle
to deduce from observations the power spectrum of the curvature
perturbation at Hubble-exit during two-field slow-roll inflation
\cite{Wands02}:
\begin{equation}
{\cal P}_\R|_* = {\cal P}_\R  \sin^2\Theta \,.
\end{equation}

The scale-dependence of the resulting scalar power spectra
depends both
upon the scale-dependence of the initial power spectra and of the
transfer coefficients.
The spectral tilts are given from Eqs.~(\ref{PR}--\ref{CRS})
by
\begin{eqnarray}
 \label{gentilt}
\Delta n_\R &=& \Delta n_\R|_* + H_*^{-1} (\partial{T}_{\R\S}/\partial t_*)
\sin 2\Theta \,,
 \nonumber\\
\Delta n_\S &=& \Delta n_\R|_* + 2 H_*^{-1} (\partial\ln{T}_{\S\S}/\partial
t_*)
\,,\\
\Delta n_{\cal C} &=& \Delta n_\R|_* + H_*^{-1} \left[
(\partial{T}_{\R\S}/\partial t_*) \tan\Theta +
(\partial\ln{T}_{\S\S}/\partial t_*) \right] \,,
 \nonumber
\end{eqnarray}
where we have used Eq.~(\ref{defTRS}) to eliminate $T_{\R\S}$ in
favour of the observable correlation angle $\Theta$.
Substituting Eq.~(\ref{nR*}) for the tilt at Hubble-exit, and
Eqs.~(\ref{dTdlnk}) for the scale-dependence of
the transfer functions, we obtain \cite{Wands02}
\begin{eqnarray}
\label{srtilts}
\Delta n_{\R} &\simeq& -(6-4\cos^2\Theta) \epsilon \nonumber\\
&&  + 2\left( \eta_{\sigma\sigma}\sin^2\Theta + 2\eta_{\sigma
 s}\sin\Theta\cos\Theta + \eta_{ss}\cos^2\Theta \right)
\,,\nonumber\\
\Delta n_\S &\simeq& -2\epsilon + 2\eta_{ss} \,,\\
\Delta n_{\cal C} &\simeq& -2\epsilon + 2\eta_{ss} + 2\eta_{\sigma
  s}\tan\Theta
 \nonumber\,.
\end{eqnarray}

Although the overall amplitude of the transfer functions are dependent
upon the evolution after Hubble-exit and through reheating into the
radiation era, the spectral tilts can be expressed solely in terms of
the slow-roll parameters at Hubble-exit during inflation and the
correlation angle, $\Theta$, which can in principle be
observed.

If the primordial curvature perturbation results solely from the
adiabatic inflaton field fluctuations during inflation then we have
$T_{\R\S}=0$ in Eq.~(\ref{PR}) and hence $\cos\Theta=0$ in
Eqs.~(\ref{srtilts}), which yields the standard single field result
\begin{equation}
 \Delta n_\R \simeq -6\epsilon +2\eta_{\sigma\sigma} \,.
\end{equation}
Any residual isocurvature perturbations must be uncorrelated with the
adiabatic curvature perturbation (at first-order in slow-roll) with
spectral index
\begin{equation}
\Delta n_\S \simeq - 2\epsilon + 2\eta_{ss} \,.
\end{equation}
%In single field inflation, by virtue of the adiabaticity of the
%large-scale perturbations (along a unique trajectory in phase space),
%the different matter components after inflation must all inherit the same
%curvature perturbation: $\zeta_\gamma=\zeta_\nu=\zeta_B=\zeta_{\rm
%c}$. The perturbations on large-scales remain adiabatic ($S_i=0$) with
%the total primordial perturbation given by the inflaton field
%fluctuations during inflation, Eq.~(\ref{Pzeta-inf}).

On the other hand, if the observed primordial curvature perturbation
is produced due to some entropy field fluctuations during inflation,
we have $T_{\R\S}\gg1$ and $\sin\Theta\simeq0$. In a two-field
inflation model any residual primordial isocurvature perturbations
will then be completely correlated (or anti-correlated) with the
primordial curvature perturbation and we have
\begin{equation}
\Delta n_\R \simeq \Delta n_{\cal C} \simeq \Delta n_\S \simeq
 - 2\epsilon + 2\eta_{ss} \,.
\end{equation}

The gravitational wave power spectrum is frozen-in
on large scales, independent of the scalar perturbations,
and hence
\begin{equation}
{\cal P}_{\rm T} = {\cal P}_{\rm T}|_* \,.
\end{equation}
Thus we can derive a modified consistency relation
\cite{LLKCreview} between observables applicable in the case of
two-field slow-roll inflation:
\begin{equation}
r=\frac{{\cal P}_{\rm T}}
{{\cal P}_\R} \simeq -8 \Delta n_{\rm T} \sin^2\Theta  \,.
\end{equation}
This relation was first obtained in Ref.~\cite{Bartolo01p2} at the
end of two-field inflation, and verified in Ref.~\cite{Tsuji03} for
slow-roll models. But it was realised in Ref.~\cite{Wands02} that
this relation also applies to the primordial perturbation spectra in
the radiation era long after two-field slow-roll inflation has ended
and hence may be tested observationally.

More generally, if there is any additional source of the scalar
curvature perturbation, such as additional scalar fields during
inflation, then this could give an additional contribution to the
primordial scalar curvature spectrum without affecting the
gravitational waves, and hence the more general result is the
inequality \cite{Sasaki95}:
\begin{equation}
r \leq -8 \Delta n_{\rm T} \sin^2\Theta \,.
\end{equation}
This leads to a fundamental difference when interpretting the
observational constraints on the amplitude of primordial tensor
perturbations in multiple inflation models.  In single field
inflation, observations directly constrain $r=[{\cal P}_{\rm T}/{\cal P}_\R]_*$ and hence, from
Eqs.~(\ref{Rstar}) and (\ref{Tstar}), the slow-roll parameter $\epsilon$. However in
multiple field inflation, non-adiabatic perturbations can enhance the
power of scalar perturbations after Hubble exit and hence
observational constraints on the amplitude of primordial tensor
perturbations do not directly constrain the slow-roll parameter
$\epsilon$.

Current CMB data alone require $r<0.55$ (assuming power-law primordial
spectra) \cite{Spergel06} which in single-field models is interpretted
as requiring $\epsilon<0.04$. But in multiple field models $\epsilon$
could be larger if the primordial density perturbation comes from
non-adiabatic perturbations during inflation.

\section{Non-Gaussianity}
%\section{$\delta N$ formalism}

A powerful technique to calculate the primordial curvature
perturbation resulting from many inflation models, including
multi-field models, is to note that the curvature perturbation
$\zeta$ defined in Eq.~(\ref{defzeta}) can be interpreted as a
perturbation in the local expansion \cite{Starobinsky,Sasaki95,Lyth05}
\begin{equation}
 \label{deltaN}
\zeta = \delta N \,,
\end{equation}
where $\delta N$ is the perturbed expansion to uniform-density
hypersurfaces with respect to spatially flat hypersurfaces, which is
given to first-order by
\begin{equation}
 \label{linearzetarho}
\zeta = - H \frac{\delta\rho_\psi}{\dot\rho} \,,
\end{equation}
where $\delta\rho_\psi$ must be evaluated on spatially flat ($\psi=0$)
hypersurfaces.

An important simplification arises on large scales where anisotropy
and spatial gradients can be neglected, and the local density,
expansion, etc, obeys the same evolution equations as in a
homogeneous FRW universe
\cite{Sasaki95,Sasaki98,Wands00,Lyth03p1,Rigopoulos03,Lyth05}. Thus
we can use the homogeneous FRW solutions to describe the local
evolution, which is known as the ``separate universe'' approach
\cite{Sasaki95,Sasaki98,Wands00,Rigopoulos03}. In particular we can
evaluate the perturbed expansion in different parts of the universe
resulting from different initial values for the fields during
inflation using the homogeneous background solutions
\cite{Sasaki95}. The integrated expansion from some initial
spatially flat hypersurface up to a late-time fixed density
hypersurface, say at the epoch of primordial nucleosynthesis, is
some function of the field values on the initial hypersurface,
$N(\varphi_I|_\psi)$. The resulting primordial curvature
perturbation on the uniform-density hypersurface is then
\begin{equation}
 \label{lineardeltaN}
\zeta = \sum_I N_{I} \delta\varphi_{I\psi} \,,
\end{equation}
where $N_I\equiv \partial N/\partial \varphi_I$ and
$\delta\varphi_{I\psi}$ is the field perturbation on some initial
spatially-flat hypersurfaces during inflation (\ref{eq:defdphipsi}).
In particular the power spectrum for the primordial density
perturbation in a multi-field inflation can be written (at leading
order) in terms of the field perturbations after Hubble-exit as
\begin{equation}
{\cal P}_\zeta = \sum_I N_I^2 {\cal
P}_{\delta\varphi_{I\psi}} \,.
\end{equation}

%\subsection{Non-linear}

This approach is readily extended to estimate the non-linear effect
of field perturbations on the metric perturbations
\cite{Sasaki98,Lyth03p1,Lyth05}.
We can take Eq.~(\ref{deltaN}) as our definition of the non-linear
primordial curvature perturbation, $\zeta$, so that in the radiation
dominated era the non-linear extension of Eq.~(\ref{linearzetarho})
is given by \cite{Lyth05}
\begin{equation}
 \label{defzetarho}
 \zeta = \frac14 \ln \left( \frac{\tilde\rho}{\rho} \right)_\psi \,,
\end{equation}
where $\tilde\rho(t,{\bf x})$ is the perturbed (inhomogeneous)
density evaluated on a spatially flat hypersurface and $\rho(t)$ is
the background (homogeneous) density. This non-linear curvature
perturbation as a function of the initial field fluctuations can
simply be expanded as a Taylor expansion
\cite{Rodriguez05,Seery05b,Byrnes06,Seery06b}
\begin{equation}
 \label{Taylor}
\zeta \simeq \sum_I N_{I} \delta\varphi_{I\psi} + \frac12
\sum_{I,J} N_{IJ} \delta\varphi_{I\psi} \delta\varphi_{J\psi}
+\frac16\sum_{I,J,K} N_{IJK} \delta\varphi_{I\psi} \delta\varphi_{J\psi} \delta\varphi_{K\psi}
+ \ldots \,.
\end{equation}
where we now identify (\ref{lineardeltaN}) as the leading-order
term.

We expect the field perturbations at Hubble-exit to be close to
Gaussian for weakly coupled scalar fields during inflation
\cite{Maldacena02,Seery05a,RS05,Seery05b,Seery06a}.  In this case the
bispectrum of the primordial curvature perturbation at leading
(fourth) order, can be written using the $\delta N$-formalism, as
\cite{Komatsu01,Rodriguez05}
\bea\label{fNLmultifielddefn} B_\zeta(\bkone,\bktwo,\bkthree) &=&
 \frac65 \fNL \left[ P_\zeta(k_1) P_\zeta(k_2) + P_\zeta(k_2)
P_\zeta(k_3) + P_\zeta(k_3) P_\zeta(k_1) \right] \,. \eea
where $P_\zeta(k)=2\pi^2{\cal P}_\zeta(k)/k^3$, and
the dimensionless non-linearity parameter is given by \cite{Rodriguez05}
\begin{equation}
 \label{fNLmultifield}
 \fNL = \frac{5}{6} \frac{N_A N_B N^{AB}}{\left(N_C N^C\right)^2} \,.
\end{equation}
Similarly, the connected part of the trispectrum in this case can be written as
\cite{Seery06b,Byrnes06}
\bea
 \label{tauNLgNLdefn}
 T_\zeta (\bkone,\bktwo,\bkthree,\bkfour) &=&
\tau_{NL}\left[P_\zeta(|{\bf k}_1+{\bf k}_3|)P_\zeta(k_3)P_\zeta(k_4)+(11\,\,\rm{perms})\right] \nonumber \\
&&+\frac{54}{25}g_{NL}\left[P_\zeta(k_2)P_\zeta(k_3)P_\zeta(k_4)+(3\,\,\rm{perms})\right]\,.
\eea
where
% $k_{ij}=|{\bf k}_i-{\bf k}_j|$,
%and
%
\bea
 \label{tauNLmultifield}
  \tau_{NL}&=&\frac{N_{AB}N^{AC}N^BN_C}{(N_DN^D)^3}\,, \\
 \label{gNLmultifield}
  g_{NL}&=&\frac{25}{54}\frac{N_{ABC}N^AN^BN^C}{(N_DN^D)^3}\,.
 \eea
The expression for $\tau_{NL}$ was first given in
\cite{Alabidi:2005qi}. Note that we have factored out products in the
trispectrum with different $k$ dependence in order to define the two
$k$ independent non-linearity parameters $\tau_{NL}$ and
$g_{NL}$. This gives the possibility that observations may be able to
distinguish between the two parameters \cite{Okamoto:2002ik}.

In many cases there is single direction in field-space, $\chi$,
which is responsible for perturbing the local expansion, $N(\chi)$,
and hence generating the primordial curvature perturbation
(\ref{Taylor}). For example this would be the inflaton field in
single field models of inflation, or it could be the late-decaying
scalar field in the curvaton scenario
\cite{Enqvist02,Lyth02,Moroi01} as will be discussed in the next
section.
In this case the curvature perturbation (\ref{Taylor}) is given by
\begin{equation}
 \label{localN}
\zeta \simeq
 N' \delta\chi_{\psi}
 + \frac12 N'' \delta\chi^2_{\psi}
 + \frac16 N''' \delta\chi^3_{\psi}
+ \ldots \,,
\end{equation}
and the non-Gaussianity of the primordial perturbation has the
simplest ``local'' form
\begin{equation}
 \label{localzeta}
\zeta = \zeta_1 + \frac35f_{NL}\zeta_1^2 + \frac{9}{25}g_{NL}\zeta_1^3 +\ldots
\end{equation}
where $\zeta_1=N'\delta\chi_\psi$ is the leading-order Gaussian
curvature perturbation and the non-linearity parameters $f_{NL}$ and
$g_{NL}$, are given by \cite{Rodriguez05,Sasaki:2006kq}
\bea
 \label{fNL1field}
  f_{NL}&=&\frac56\frac{N''}{(N')^2}\,, \\
 \label{gNL1field}
  g_{NL}&=&\frac{25}{54}\frac{N'''}{(N')^3}\,,
\eea
The primordial bispectrum and trispectrum are then given by
Eqs.~(\ref{fNLmultifielddefn}) and~(\ref{tauNLgNLdefn}), where the
non-linearity parameters $f_{NL}$ and $g_{NL}$, given in
Eqs.~(\ref{fNLmultifield}) and~(\ref{gNLmultifield}), reduce to
Eqs.~(\ref{fNL1field}) and~(\ref{gNL1field}) respectively,
and $\tau_{NL}$ given in Eq.(\ref{tauNLmultifield}) reduces to
\bea
%\label{fNL1field} f_{NL}&=&\frac56\frac{N''}{(N')^2}\,, \\
\label{tauNL1field}
 \tau_{NL}&=&\frac{(N'')^2}{(N')^4}=\frac{36}{25}f_{NL}^2\,.
% \\
%\label{gNL1field} g_{NL}&=&\frac{25}{54}\frac{N'''}{(N')^3}\,,
\eea
Thus $\tau_{NL}$ is proportional to $f_{NL}^2$ (first shown in
\cite{Okamoto:2002ik} using the Bardeen potential, and in
\cite{Rodriguez05} using this notation). However the trispectrum could
be large even when the bispectrum is small because of the $g_{NL}$
term \cite{Okamoto:2002ik,Sasaki:2006kq}.

In the case of where the primordial curvature perturbation is
generated solely by adiabatic fluctuations in the inflaton field,
$\sigma$, the curvature perturbation is non-linearly conserved on
large scales \cite{Lyth03,Lyth05,Langlois:2005ii} and we can
calculate $N'$, $N''$, $N'''$, etc, at Hubble-exit. In terms of the
slow-roll parameters, we find
\bea N'&=& \frac{{H}}{\dot{{\varphi}}}
\simeq \frac{1}{\sqrt{2}}\frac{1}{\Mp}\frac{1}{\sqrt{\epsilon}}\sim\mathcal{O}\left(\epsilon^{-\frac12}\right)\,,  \\
N''&\simeq&-\frac12\frac{1}{\Mp^2}\frac{1}{\epsilon}(\eta_{\sigma\sigma}-2\epsilon)\sim\mathcal{O}\left(1\right)\,,  \\
N'''&\simeq&\frac{1}{\sqrt{2}}\frac{1}{\Mp^3}\frac{1}{\epsilon\sqrt{\epsilon}}\left(\epsilon\eta_{\sigma\sigma}-\eta_{\sigma\sigma}^2+\frac12\xi_\sigma^2\right)
\sim\mathcal{O}(\epsilon^{\frac12})\,, \eea
where we have used the reduced Planck mass $\Mp^2=(8\pi G)^{-1}$ and
introduced the second-order slow-roll parameter
$\xi_\sigma^2=\Mp^4V_\sigma V_{\sigma\sigma\sigma}/V^2$.
Hence the non-linearity parameters for single field inflation,
(\ref{fNL1field}) and~(\ref{gNL1field}), are given by
\bea
 f_{NL}&=&\frac56(\eta_{\sigma\sigma}-2\epsilon)\,, \\
% \tau_{NL}&=&(\eta_{\sigma\sigma}-2\epsilon)^2\,, \\
 g_{NL}&=&\frac{25}{54}\left(2\epsilon\eta_{\sigma\sigma}-2\eta_{\sigma\sigma}^2+
 \xi_\sigma^2\right)\,. \eea
with $\tau_{NL}$ given by Eq.~(\ref{tauNL1field}). Although there
are additional contributions to the primordial bispectrum and
trispectrum coming from the intrinsic non-Gaussianity of the field
perturbations at Hubble-exit, these are also suppressed by slow-roll
parameters in slow-roll inflation. Thus the primordial
non-Gaussianity is likely to be too small to ever be observed in the
conventional inflaton scenario of single-field slow-roll inflation.
Indeed any detection of primordial non-Gaussianity $f_{NL}>1$ would
appear to rule out this inflaton scenario.

However significant non-Gaussianity can be generated due to
non-adiabatic field fluctuations. Thus far it has proved difficult to
generate significant non-Gaussianity in the curvature perturbation
during slow-roll inflation, even in multiple field models. But
detectable non-Gaussianity can be produced when the curvature
perturbation is generated from isocurvature field perturbations at the
end of inflation \cite{BernardeauUzan,LythRiotto}, during
inhomogeneous reheating \cite{Zaldarriaga,Kolb,Byrnes05}, or
after inflation in the curvaton model, which I will discuss next.

\section{Curvaton scenario}

Consider a light, weakly-coupled scalar field, $\chi$, that decays
some time after inflation has ended. There are many such scalar
degrees of freedom in supersymmetric theories and if they are too
weakly coupled, and their lifetime is too long, this leads to the
``Polonyi problem'' \cite{Linde90}. Assuming the field is displaced from the
minimum of its effective potential at the end of inflation, the
field evolves little until the Hubble rate drops below its effective
mass. Then it oscillates, with a time-averaged equation of state of
a pressureless fluid, $P_\chi=0$, (or, equivalently, a collection of
non-relativistic particles) and will eventually come to dominate the
energy density of the Universe. To avoid disrupting the standard
``hot big bang'' model and in particular to preserve the successful
radiation-dominated model of primordial nucleosythesis, we require
that such fields decay into radiation before $t\sim 1$~second. For a
weakly-coupled field that decays with only gravitational strength,
$\Gamma\sim m_\chi^3/M_P^2$, this requires $m_\chi>100$~TeV.

But there is a further important feature of late-decaying scalar
fields that has only recently received serious consideration. If the
field is inhomogeneous then it could lead to an inhomogeneous
radiation density after it decays
\cite{Mollerach90,LindeMukhanov96}. This is the basis of the
curvaton scenario \cite{Enqvist02,Lyth02,Moroi01}.

If this field is light ($m<H$) during inflation then small-scale
quantum fluctuations will lead to a spectrum of large-scale
perturbations, whose initial amplitude at Hubble-exit is given (at
leading order in slow-roll) by Eq.~(\ref{H2pi}). When the Hubble
rate drops and the field begins oscillating after inflation, this
leads to a first-order density perturbation in the $\chi$-field:
\begin{equation}
 \label{defzetachi}
\zeta_\chi = -\psi + \frac{\delta\rho_\chi}{3\rho_\chi} \,.
\end{equation}
where $\rho_\chi=m_\chi^2\chi^2/2$. $\zeta_\chi$ remains constant
for the oscillating curvaton field on large scales, so long as we
can neglect its energy loss due to decay. Using Eq.~(\ref{H2pi}) for the
field fluctuations at Hubble-exit and neglecting any non-linear
evolution of the $\chi$-field after inflation (consistent with our
assumption that the field is weakly coupled), we have
\begin{equation}
 \label{Pzetachi}
{\cal P}_{\zeta_\chi} \simeq \left( \frac{H}{6\pi\chi}
\right)^2_{k=aH} \,.
\end{equation}
The total density perturbation (\ref{defzeta}), considering
radiation, $\gamma$, and the curvaton, $\chi$, is given by \cite{Lyth02}
\begin{equation}
\zeta =
\frac{4\rho_\gamma\zeta_\gamma+3\rho_\chi\zeta_\chi}{4\rho_\gamma+3\rho_\chi}
 \,.
\end{equation}
Thus if the radiation generated by the decay of the inflaton at the
end of inflation is unperturbed (${\cal P}_{\zeta_\gamma}^{1/2}\ll
10^{-5}$) the total curvature perturbation grows as the density of
the $\chi$-field grows relative to the radiation:
$\zeta\sim\Omega_\chi\zeta_\chi$.

Ultimately the $\chi$-field must decay (when $H\sim\Gamma$) and
transfer its energy density and, crucially, its perturbation to the
radiation and/or other matter fields. In the simplest case that the
non-relativistic $\chi$-field decays directly to radiation a full
analysis \cite{MWU,Gupta} of the coupled evolution equation gives
the primordial radiation perturbation (after the decay)
\begin{equation}
 \label{defrp}
\zeta = r_\chi(p) \zeta_\chi \,,
\end{equation}
where $p\equiv [\Omega_\chi/(\Gamma/H)^{1/2}]_{\rm initial}$ is a
dimensionless parameter which determines the maximum value of
$\Omega_\chi$ before it decays, and empirically we find \cite{Gupta}
\begin{equation}
r_\chi(p) \simeq 1- \left( 1+ \frac{0.924}{1.24}p \right)^{-1.24}  \,.
\end{equation}
For $p\gg1$ the $\chi$-field dominates the total energy density
before it decays and $r_\chi\sim1$, while for $p\ll1$ we have
$r_\chi\sim0.924p\ll1$.

Finally combining Eqs.~(\ref{Pzetachi}) and (\ref{defrp}) we
have
\begin{equation}
 \label{Pzetagamma}
{\cal P}_{\zeta} \simeq r_\chi^{\,2}(p) \left( \frac{H}{6\pi\chi}
 \right)^2_{k=aH} \,.
\end{equation}
Note that the primordial curvature perturbation tends to have less
power on small scales due to the decreasing Hubble rate at Hubble
exit in Eq.~(\ref{Pzetagamma}), but can also have more power on
small scales due to the decreasing $\chi$, for a positive effective
mass-squared, during inflation. In terms of slow-roll parameters the
actual tilt is given by Eq.~(\ref{srtilts}) when $\sin\Theta=0$
\begin{equation}
 \Delta n_\R \simeq -2\epsilon + 2\eta_{\chi\chi} \,.
\end{equation}
In the extreme slow-roll limit the spectrum becomes scale-invariant,
as in the inflaton scenario.

In contrast to the inflaton scenario the final density perturbation in
the curvaton scenario is a very much dependent upon the physics after
the field perturbation exited the Hubble scale during inflation. For
instance, if the curvaton lifetime is too short then it will decay
before it can significantly perturb the total energy density and
${\cal P}_{\zeta_\gamma}^{1/2}\ll 10^{-5}$. The observational
constraint on the amplitude of the primordial perturbations gives a
single constraint upon both the initial fluctuations during inflation
and the post-inflationary decay time. This is in contrast to the
inflaton scenario where the primordial perturbation gives a direct
window onto the dynamics of inflation, independently of the physics at
lower energies. In the curvaton scenario there is the possibility of
connecting the generation of primordial perturbations to other aspects
of cosmological physics. For instance, it may be possible to identify
the curvaton with fields whose late-decay is responsible for the
origin of the baryon asymmetry in the universe, in particular with
sneutrino models of leptogenesis (in which an initial lepton asymmetry
is converted into a baryon asymmetry at the electroweak transition)
\cite{sneutrino}.

The curvaton scenario has re-invigorated attempts to embed models of
inflation in the very early universe within minimal supersymmetric
models of particle physics constrained by experiment \cite{mssm}. It
may be possible that the inflaton field driving inflation can be
completely decoupled from visible matter if the dominant radiation
in the universe today comes from the curvaton decay rather than
reheating at the end of inflation. Indeed the universe need not be
radiation-dominated at all until the curvaton decays if instead the
inflaton fast-rolls at the end of inflation.

The curvaton offers a new range of theoretical possibilities, but
ultimately we will require observational and/or experimental
predictions to decide whether the curvaton, inflaton or some othe
field generated the primordial perturbation. I will discuss
observational predictions of the curvaton scenario in the following
subsection.

\subsection{Non-Gaussianity}

The best way to distinguish between different scenarios for the
origin of structure could be the statistical properties of the
primordial density perturbation.
Primordial density perturbations in the curvaton scenario originate
from the small-scale vacuum fluctuations of the weakly interacting
curvaton field during inflation, which can be described on
super-Hubble scales by a Gaussian random field.  Thus deviations from
Gaussianity in the primordial bispectrum and connected trispectrum can
be parameterised by the dimensionless parameters $f_{NL}$ and $g_{NL}$
defined in Eqs.~(\ref{fNL1field}) and~(\ref{gNL1field}).

When the curvaton field begins oscillating about a quadratic minimum
of its potential we have $\rho_\chi=m_\chi^2\chi^2/2$, and the time-averaged
equation of state becomes, $P_\chi=0$.
The non-linear generalisation of the primordial curvature perturbation
(\ref{defzetaI}) on hypersurfaces of uniform-curvaton density is then
\begin{equation}
 \label{NLzetachi}
 \zeta_\chi = \frac13 \ln \left( \frac{\tilde\rho_{\chi}}{\rho_\chi}
 \right)_\psi \,,
\end{equation}
where we distinguish here between the inhomogeneous density
$\tilde\rho_{\chi}$ on spatially flat hypersurfaces and the average
density $\rho_\chi$. Given that $\rho_\chi\propto\chi^2$ we thus
have
\begin{equation}
 \zeta_\chi = \frac13 \ln \left( 1 +
 \frac{2\chi\delta\chi+\delta\chi^2}{\chi^2} \right)_\psi \,.
\end{equation}
This gives the full probability distribution function for
$\zeta_\chi$ for Gaussian field perturbations $\delta\chi$.
Expanding to first-order we obtain $\zeta_{\chi
1}=2\delta\chi_\psi/3\chi$, and then to second- and third-order we
obtain by analogy with Eq.~(\ref{localzeta}) the non-linearity
parameters for the curvaton density perturbation
\begin{equation}
 \zeta_\chi \simeq
 \zeta_{\chi1}
 + \frac35 f_{NL}^\chi \zeta_{\chi1}^2
 + \frac{9}{25} g_{NL}^\chi \zeta_{\chi1}^3
+ \ldots \,,
\end{equation}
where \cite{Sasaki:2006kq}
\begin{eqnarray}
 \label{fNLchi}
f_{NL}^\chi = - \frac54 \,,\\
 \label{gNLchi}
g_{NL}^\chi = \frac{25}{12} \,.
\end{eqnarray}
If the curvaton dominates the total energy density before it decays
into radiation, then this is the curvature perturbation, and
specifically the non-linearity parameters, inherited by the primordial
radiation density. Although not suppressed by slow-roll parameters,
this non-Gaussianity is still smaller than the best upper limits
expected from the Planck satellite \cite{Komatsu01}.

On the other hand if the curvaton decays before it dominates over the
energy density of the existing radiation, so the transfer function
$r_\chi(p)\ll1$ in Eq.~(\ref{defrp}), then the curvaton may lead to a
large and detectable non-Gaussianity in the radiation density after it
decays. Assuming the sudden decay of the curvaton on the $H=\Gamma$
uniform-density hypersurface leads to a non-linear relation between
the local curvaton density and the radiation density before and after
the decay \cite{Sasaki:2006kq}
\begin{equation}
 \label{NLzetas}
 \rho_\gamma e^{-4\zeta} + \rho_\chi e^{3(\zeta_\chi-\zeta)} =
 \rho_\gamma + \rho_\chi \,.
\end{equation}
Expanding this term by term yields
\cite{Bartolo04cur,Rodriguez05,Sasaki:2006kq}
\begin{eqnarray}
f_{NL} &=& \frac{5}{4r_\chi} - \frac53 - \frac{5r_\chi}{6} \,,\\
g_{NL} &=& - \frac{25}{6r_\chi} +\frac{25}{108} + \frac{125r_\chi}{27} +
\frac{25r_\chi^2}{18} \,.
\end{eqnarray}
These reduce to Eqs.~(\ref{fNLchi}) and~(\ref{gNLchi}) as $r_\chi\to1$,
but become large for $r_\chi\ll1$.
These analytic results rely on the sudden decay approximation but have
been tested against numerical solutions \cite{ML06,Sasaki:2006kq} and
give an excellent approximation for both $r_\chi\ll1$ and $r_\chi\simeq1$.

More generally one can use Eqs.~(\ref{NLzetachi})
  and~(\ref{NLzetas}), or use Eq.~(\ref{deltaN}) and solve the
  non-linear, but homogeneous equations of motion to determine
  $N(\chi)$ to give the full probability distribution function for the
  primordial curvature perturbation $\zeta$ \cite{Sasaki:2006kq}.

Current bounds from the WMAP satellite require $-54<f_{NL}<114$ at
the 95\% confidence limit \cite{Spergel06}, and hence require
$r_\chi>0.011$. But future cosmic microwave background (CMB) experiments
such as Planck could detect $f_{NL}$ as small as around $5$
\cite{Komatsu01}, and it has even been suggested that it might one
day be possible to constrain $f_{NL}\sim 0.01$ \cite{Cooray06}.

\subsection{Residual isocurvature perturbations}

In the curvaton scenario the initial curvaton perturbation is a
non-adiabatic perturbation and hence can in principle leave behind a
residual non-adiabatic component. Perturbations in this one field,
would be responsible for both the total primordial density
perturbation and any isocurvature mode and hence there is the clear
prediction that the two should be completely correlated,
corresponding to $\cos\Theta=\pm1$ in Eq.~(\ref{defDelta})
%, or $A_r/A_s\to0$ in Eq.~(\ref{defArAs})
and $\Delta n_{\cal R}-1=\Delta n_{\cal S}=\Delta n_{\cal C}$ in Eqs.~(\ref{srtilts}).

Using $\zeta_I$ defined in Eq.~(\ref{defzetaI}) for different matter
components it is easy to see how the curvaton could leave residual
isocurvature perturbations after the curvaton decays. If any fluid
has decoupled before the curvaton contributes significantly to the
total energy density that fluid remains unperturbed with
$\zeta_I\simeq0$, whereas after the curvaton decays into radiation
the photons perturbation is given by (\ref{defrp}). Thus a residual
isocurvature perturbation (\ref{defSI}) is left
\begin{equation}
S_I = -3 \zeta \,,
\end{equation}
which remains constant for decoupled perfect fluids on large scales.

The observational bound on isocurvature matter perturbations
completely correlated with the photon perturbation, is
\cite{Lewis06}
\begin{equation}
 \label{Bbound}
-0.42 <\frac{S_B+(\rho_{\rm cdm}/\rho_B)S_{\rm cdm}} {\zeta_\gamma}
< 0.25 \,.
\end{equation}
In particular if the baryon asymmetry is generated while the total
density perturbation is still negligible then the residual baryon
isocurvature perturbation, $S_B=-3\zeta_\gamma$ would be much larger
than the observational bound and such models are thus ruled out. The
observational bound on CDM isocurvature perturbations are stronger
by the factor $\rho_{\rm cdm}/\rho_B$ \cite{Gordon02} although CDM
is usually assumed to decouple relatively late.

An interesting amplitude of residual isocurvature perturbations
might be realised if the decay of the curvaton itself is the
non-equilibrium event that generates the baryon asymmetry. In this
case the net baryon number density directly inherits the
perturbation $\zeta_B=\zeta_\chi$ while the photon perturbation
$\zeta_\gamma\leq\zeta_\chi$ may be diluted by pre-existing
radiation and is given by Eq.~(\ref{defrp}). Note that so long as
the net baryon number is locally conserved it defines a conserved
perturbation on large scales, even though it may still be
interacting with other fluids and fields \cite{Lyth03}. Hence the
primordial baryon isocurvature perturbation (\ref{defSB}) in this
case is given by
\begin{equation}
S_B = 3 (1-r_\chi) \zeta_\chi = \frac{3(1-r_\chi)}{r_\chi} \zeta_\gamma \,.
\end{equation}
Thus the observational bound (\ref{Bbound}) requires $r_\chi>0.92$
if the baryon asymmetry is generated by the curvaton decay.

There is no lower bound on the predicted amplitude of residual
non-adiabatic modes and, although the detection of completely
correlated isocurvature perturbations would give strong support to the
curvaton scenario, the non-detection of primordial isocurvature
density perturbations cannot be used to rule out the curvaton
scenario. In particular, if the curvaton decays at sufficiently high
temperature and all the particles produced relax to a thermal
equilibrium abundance, characterised by a common temperature (and
vanishing chemical potentials) then no residual isocurvature
perturbations survive. In full thermal equilibrium there is a unique
attractor trajectory in phase-space and only adiabatic perturbations
(along this trajectory) survive on large scales.

\section{Conclusions}

Inflation offers a beautifully simple origin for structure in our
Universe. The zero-point fluctuations of the quantum vacuum state on
sub-atomic scales are swept up by the accelerated expansion to
astronomical scales, seeding an almost Gaussian distribution of
primordial density perturbations. The large scale structure of our
Universe can then form simply due to the gravitational instability
of overdense regions.

Astronomical observations over recent years have given strong
support to this simple picture. But increasingly precise
astronomical data will increasingly allow us to probe not only the
parameters of what has become the standard cosmological model, but
also to probe the nature of the primordial perturbations from which
structure formed. Any evidence of primordial gravitational waves,
primordial isocurvature fluctuations, and/or non-Gaussianity of the
primordial perturbations could provide valuable information about
the inflationary dynamics that preceded the hot big bang.

Single-field slow-roll inflation predicts adiabatic density
perturbations with negligible non-Gaussianity, but could produce a
gravitational wave background which could be detected by upcoming
CMB experiments.

On the other hand multiple field inflation can lead to a wider range
of possibilities which could be distinguished by observations. A
spectrum of non-adiabatic field fluctuations on large scales during
inflation could leave residual isocurvature perturbations after
inflation, which can be correlated with the primordial curvature
perturbation, and can give rise to a detectable level of
non-Gaussianity. In simple models, such as the curvaton scenario,
where the primordial curvature perturbations originate from almost
Gaussian fluctuations in a single scalar field, any residual
isocurvature perturbations are expected to be completely correlated
with the curvature perturbation and the non-Gaussianity is of a
specific ``local'' form.

After 25 years studying inflation, we may for the first time have
evidence of a weak scale-dependence of the power spectrum of the
primordial curvature perturbation \cite{Spergel06} which would begin
to reveal the slow-roll dynamics during inflation.
Primordial perturbations may have much more to tell us about
the physics of inflation in the future.

\section*{Acknowledgements}

I am grateful to the organisers of the Inflation +25 conference and
workshop for their hospitality, and my numerous collaborators for
their contributions to the work which I have reported
here. I am also grateful to Chris Byrnes, Karim Malik and Jussi
V\"aliviita for their comments on this article.

%
%
% BibTeX users please use
% \bibliographystyle{}
% \bibliography{}
%
% Non-BibTeX users please follow the syntax
% the syntax of "referenc.tex" for your own citations
%\input{wands-referenc}
%%%%%%%%%%%%%%%%%%%%%%%%%%%%%%%%%%%%%%%%%%%%%%%%%%%%%%%%%%%%%%%%%%%%%%  }

%%%%%%%%%%%%%%%%%%%%%%%% referenc.tex %%%%%%%%%%%%%%%%%%%%%%%%%%%%%%
% sample references
% "physics"
%
% Use this file as a template for your own input.
%
%%%%%%%%%%%%%%%%%%%%%%%% Springer-Verlag %%%%%%%%%%%%%%%%%%%%%%%%%%

%
% BibTeX users please use
% \bibliographystyle{}
% \bibliography{}

\begin{thebibliography}{99.}
% Use the following syntax and markup for your references
%
% Monographs
%\bibitem{monograph} H. Ibach, H. L\"uth: \textit{Solid-State
%Physics}, 2nd edn (Springer, Berlin Heidelberg New York 1996) pp 45--56

% Contributed Works
%\bibitem{contribution} D.M. MacKay: Visual stability and voluntary eye
%movements. In: \textit{Handbook of Sensory Physiology}, vol 3, ed by R.
%Jung, D.M. MacKay (Springer, Berlin Heidelberg New York 1973) pp
%307--331

% Journal
%\bibitem{journal} S. Preuss, A. Demchuk Jr, M. Stuke et al: Appl. Phys.
%A \textbf{61}, 33 (1995)

% Theses
%\bibitem{thesis} D.W.  Ross: Lysosomes and storage diseases. MA
%Thesis, Columbia University, New York (1977)

\bibitem{RMP}
  B.~A.~Bassett, S.~Tsujikawa and D.~Wands,
  %``Inflation dynamics and reheating,''
  Rev.\ Mod.\ Phys.\  {\bf 78}, 537 (2006)
  [arXiv:astro-ph/0507632].
  %%CITATION = ASTRO-PH 0507632;%%

\bibitem{Linde}
%\bibitem{Linde:1993cn}
  A.~D.~Linde,
  %``Hybrid inflation,''
  Phys.\ Rev.\ D {\bf 49}, 748 (1994)
  [arXiv:astro-ph/9307002].
  %%CITATION = ASTRO-PH 9307002;%%

\bibitem{Copeland}
%\bibitem{Copeland:1994vg}
  E.~J.~Copeland, A.~R.~Liddle, D.~H.~Lyth, E.~D.~Stewart and D.~Wands,
  %``False vacuum inflation with Einstein gravity,''
  Phys.\ Rev.\ D {\bf 49}, 6410 (1994)
  [arXiv:astro-ph/9401011].
  %%CITATION = ASTRO-PH 9401011;%%


\bibitem{Liddle}
%\bibitem{Liddle:1998jc}
  A.~R.~Liddle, A.~Mazumdar and F.~E.~Schunck,
  %``Assisted inflation,''
  Phys.\ Rev.\ D {\bf 58}, 061301 (1998)
  [arXiv:astro-ph/9804177].
  %%CITATION = ASTRO-PH 9804177;%%


\bibitem{Malik}
%\bibitem{Malik:1998gy}
  K.~A.~Malik and D.~Wands,
  %``Dynamics of assisted inflation,''
  Phys.\ Rev.\ D {\bf 59}, 123501 (1999)
  [arXiv:astro-ph/9812204].
  %%CITATION = ASTRO-PH 9812204;%%

\bibitem{Gordon00}
C.~Gordon, D.~Wands, B.~A.~Bassett and R.~Maartens,
%``Adiabatic and entropy perturbations from inflation,''
Phys.\ Rev.\ D {\bf 63}, 023506 (2001).

\bibitem{Groot01}
S.~Groot Nibbelink and B.~J.~W.~van Tent,
%``Scalar perturbations during multiple
% field slow-roll inflation,''
Class.\ Quant.\ Grav.\  {\bf 19}, 613 (2002).

\bibitem{Rigopoulos}
G.~Rigopoulos,
%``On second order gauge invariant
%perturbations in multi-field inflationary models,''
Class.\ Quant.\ Grav.\  {\bf 21}, 1737 (2004).

\bibitem{Nflation}
  S.~Dimopoulos, S.~Kachru, J.~McGreevy and J.~G.~Wacker,
  %``N-flation,''
  arXiv:hep-th/0507205.
  %%CITATION = HEP-TH 0507205;%%

\bibitem{KO}
  P.~Kanti and K.~A.~Olive,
  %``On the realization of assisted inflation,''
  Phys.\ Rev.\ D {\bf 60}, 043502 (1999)
  [arXiv:hep-ph/9903524];
  %%CITATION = HEP-PH 9903524;%%
  %``Assisted chaotic inflation in higher dimensional theories,''
  Phys.\ Lett.\ B {\bf 464}, 192 (1999)
  [arXiv:hep-ph/9906331].
  %%CITATION = HEP-PH 9906331;%%

\bibitem{KL}
  N.~Kaloper and A.~R.~Liddle,
  %``Dynamics and perturbations in assisted chaotic inflation,''
  Phys.\ Rev.\ D {\bf 61}, 123513 (2000)
  [arXiv:hep-ph/9910499].
  %%CITATION = HEP-PH 9910499;%%

\bibitem{LythRiotto}
  D.~H.~Lyth and A.~Riotto,
  %``Particle physics models of inflation and the cosmological density
  %perturbation,''
  Phys.\ Rept.\  {\bf 314}, 1 (1999)
  [arXiv:hep-ph/9807278].
  %%CITATION = HEP-PH 9807278;%%

\bibitem{Kim}
  S.~A.~Kim and A.~R.~Liddle,
  %``Nflation: Multi-field inflationary dynamics and perturbations,''
  Phys.\ Rev.\ D {\bf 74}, 023513 (2006)
  [arXiv:astro-ph/0605604].
  %%CITATION = ASTRO-PH 0605604;%%

\bibitem{Byrnes05}
  C.~T.~Byrnes and D.~Wands,
  %``Scale-invariant perturbations from chaotic inflation,''
  Phys.\ Rev.\ D {\bf 73}, 063509 (2006)
  [arXiv:astro-ph/0512195].
  %%CITATION = ASTRO-PH 0512195;%%

\bibitem{Mukhanov85}
V.~F.~Mukhanov,
%``Gravitational Instability Of The Universe
%Filled With A Scalar Field,''
JETP Lett.\  {\bf 41}, 493 (1985)
[Pisma Zh.\ Eksp.\ Teor.\ Fiz.\  {\bf 41}, 402 (1985)].

\bibitem{Sasaki86}
M.~Sasaki,
%``Large Scale Quantum Fluctuations
% In The Inflationary Universe,''
Prog.\ Theor.\ Phys.\  {\bf 76}, 1036 (1986).

\bibitem{Bardeen83}
J.~M.~Bardeen, P.~J.~Steinhardt and M.~S.~Turner,
Phys.\ Rev.\ D {\bf 28}, 679 (1983).

\bibitem{Bardeen88}
J.~M.~Bardeen,
{\it Lectures given at 2nd Guo Shou-jing Summer School on Particle
Physics and Cosmology, Nanjing, China, Jul 1988.}

\bibitem{Martin98}
J.~Martin and D.~J.~Schwarz,
%``The influence of cosmological transitions on
%the evolution of density perturbations,''
Phys.\ Rev.\ D {\bf 57}, 3302 (1998).

\bibitem{Wands00}
D.~Wands, K.~A.~Malik, D.~H.~Lyth and A.~R.~Liddle,
%``A new approach to the evolution of cosmological
%perturbations on large scales,''
Phys.\ Rev.\ D {\bf 62}, 043527 (2000).

\bibitem{Lukash80}
V.~N.~Lukash,
%``Production Of Phonons In An Isotropic Universe,''
Sov.\ Phys.\ JETP {\bf 52}, 807 (1980).

\bibitem{Lyth85}
D.~H.~Lyth,
%``Large Scale Energy Density Perturbations And Inflation,''
Phys.\ Rev.\ D {\bf 31}, 1792 (1985).

\bibitem{Bardeen80}
J.~M.~Bardeen,
%``Gauge Invariant Cosmological Perturbations,''
Phys.\ Rev.\ D {\bf 22}, 1882 (1980).

\bibitem{Lyth03p1}
D.~H.~Lyth and D.~Wands,
%``Conserved cosmological perturbations,''
Phys.\ Rev.\ D {\bf 68}, 103515 (2003).

\bibitem{Taruya:1997iv}
  A.~Taruya and Y.~Nambu,
  %``Cosmological perturbation with two scalar fields in reheating after
  %inflation,''
  Phys.\ Lett.\ B {\bf 428}, 37 (1998)
  [arXiv:gr-qc/9709035].
  %%CITATION = GR-QC 9709035;%%

\bibitem{MW}
  K.~A.~Malik and D.~Wands,
  %``Adiabatic and entropy perturbations with interacting fluids and fields,''
  JCAP {\bf 0502}, 007 (2005)
  [arXiv:astro-ph/0411703].
  %%CITATION = ASTRO-PH 0411703;%%

\bibitem{Moodley}
  M.~Bucher, K.~Moodley and N.~Turok,
  %``Characterising the primordial cosmic perturbations using MAP and  PLANCK,''
  Phys.\ Rev.\ D {\bf 66}, 023528 (2002)
  [arXiv:astro-ph/0007360].
  %%CITATION = ASTRO-PH 0007360;%%

\bibitem{Juan}
  M.~Beltran, J.~Garcia-Bellido, J.~Lesgourgues and M.~Viel,
  %``Squeezing the window on isocurvature modes with the Lyman-alpha forest,''
  Phys.\ Rev.\ D {\bf 72}, 103515 (2005)
  [arXiv:astro-ph/0509209].
  %%CITATION = ASTRO-PH 0509209;%%

\bibitem{Bean}
  R.~Bean, J.~Dunkley and E.~Pierpaoli,
  %``Constraining Isocurvature Initial Conditions with WMAP 3-year data,''
  Phys.\ Rev.\ D {\bf 74}, 063503 (2006)
  [arXiv:astro-ph/0606685].
  %%CITATION = ASTRO-PH 0606685;%%

\bibitem{Trotta}
  R.~Trotta,
  %``Probing dark energy with future surveys,''
  arXiv:astro-ph/0607496.
  %%CITATION = ASTRO-PH 0607496;%%

\bibitem{Jussi}
  R.~Keskitalo, H.~Kurki-Suonio, V.~Muhonen and J.~Valiviita,
  %``Hints of Isocurvature Perturbations in the Cosmic Microwave Background,''
  arXiv:astro-ph/0611917.
  %%CITATION = ASTRO-PH 0611917;%%

\bibitem{DiMarco03}
F.~Di Marco, F.~Finelli and R.~Brandenberger,
%``Adiabatic and Isocurvature Perturbations for
%Multifield Generalized Einstein Models,''
Phys.\ Rev.\ D {\bf 67}, 063512 (2003).

\bibitem{Choi07}
  K.~Y.~Choi, L.~M.~H.~Hall and C.~v.~de Bruck,
  %``Spectral running and non-Gaussianity from slow-roll inflation in
  %generalised two-field models,''
  arXiv:astro-ph/0701247.
  %%CITATION = ASTRO-PH 0701247;%%

\bibitem{Mukhanov90}
V.~F.~Mukhanov, H.~A.~Feldman and R.~H.~Brandenberger,
 %``Theory Of Cosmological Perturbations. Part 1.
 %Classical Perturbations. Part
%2. Quantum Theory Of Perturbations. Part 3. Extensions,''
Phys.\ Rept.\  {\bf 215}, 203 (1992).

\bibitem{Weinberg04a}
S.~Weinberg,
%``Can non-adiabatic perturbations arise
% after single-field inflation?,''
Phys.\ Rev.\ D {\bf 70}, 043541 (2004).

\bibitem{Weinberg04b}
S.~Weinberg,
%``Must cosmological perturbations remain non-adiabatic
%after multi-field inflation?,''
Phys.\ Rev.\ D {\bf 70}, 083522 (2004).

\bibitem{Langlois99}
D.~Langlois,
%``Correlated Adiabatic And Isocurvature
%Perturbations From Double Inflation,''
Phys.\ Rev.\ D {\bf 59}, 123512 (1999).

\bibitem{Wands02}
D.~Wands, N.~Bartolo, S.~Matarrese and A.~Riotto,
%``An observational test of two-field inflation,''
Phys.\ Rev.\ D {\bf 66}, 043520 (2002).

\bibitem{LLKCreview}
J.~E.~Lidsey, A.~R.~Liddle, E.~W.~Kolb, E.~J.~Copeland,
%``Reconstructing the inflaton potential: An overview,''
Rev.\ Mod.\ Phys.\  {\bf 69}, 373 (1997).

\bibitem{Bartolo01p2}
N.~Bartolo, S.~Matarrese and A.~Riotto,
%``Adiabatic and isocurvature perturbations
%from inflation: Power spectra  and
%consistency relations,''
Phys.\ Rev.\ D {\bf 64}, 123504 (2001).

\bibitem{Tsuji03}
S.~Tsujikawa, D.~Parkinson and B.~A.~Bassett,
%``Correlation-consistency cartography
%of the double inflation landscape,''
Phys.\ Rev.\ D {\bf 67}, 083516 (2003).

\bibitem{Sasaki95}
M.~Sasaki and E.~D.~Stewart,
%``A General analytic formula for the spectral index of the density
%perturbations produced during inflation,''
Prog.\ Theor.\ Phys.\  {\bf 95}, 71 (1996).

\bibitem{Spergel06}
  D.~N.~Spergel {\it et al.},
  %``Wilkinson Microwave Anisotropy Probe (WMAP) three year results:
  %Implications for cosmology,''
  arXiv:astro-ph/0603449.
  %%CITATION = ASTRO-PH 0603449;%%

\bibitem{Starobinsky}
  A.~A.~Starobinsky,
  %``MULTICOMPONENT DE SITTER (INFLATIONARY) STAGES AND THE GENERATION OF
  %PERTURBATIONS,''
  JETP Lett.\  {\bf 42}, 152 (1985)
  [Pisma Zh.\ Eksp.\ Teor.\ Fiz.\  {\bf 42}, 124 (1985)].
  %%CITATION = JTPLA,42,152;%%

\bibitem{Sasaki98}
M.~Sasaki and T.~Tanaka,
%``Super-horizon scale dynamics of multi-scalar inflation,''
Prog.\ Theor.\ Phys.\  {\bf 99}, 763 (1998).

\bibitem{Lyth05}
D.~H.~Lyth, K.~A.~Malik and M.~Sasaki,
%``A general proof of the conservation of
% the curvature perturbation,''
JCAP {\bf 0505}, 004 (2005).

\bibitem{Rigopoulos03}
G.~I.~Rigopoulos and E.~P.~S.~Shellard,
%``The separate universe approach and the evolution of
%nonlinear  superhorizon cosmological perturbations,''
Phys.\ Rev.\ D {\bf 68}, 123518 (2003).

\bibitem{Byrnes06}
  C.~T.~Byrnes, M.~Sasaki and D.~Wands,
  %``The primordial trispectrum from inflation,''
  Phys.\ Rev.\ D {\bf 74}, 123519 (2006)
  [arXiv:astro-ph/0611075].
  %%CITATION = ASTRO-PH 0611075;%%

\bibitem{vanTent}
  B.~van Tent,
  %``Multiple-field inflation and the CMB,''
  Class.\ Quant.\ Grav.\  {\bf 21}, 349 (2004)
  [arXiv:astro-ph/0307048].
  %%CITATION = ASTRO-PH 0307048;%%

\bibitem{Byrnes:2006fr}
  C.~T.~Byrnes and D.~Wands,
  %``Curvature and isocurvature perturbations from two-field inflation in a
  %slow-roll expansion,''
  Phys.\ Rev.\ D {\bf 74}, 043529 (2006)
  [arXiv:astro-ph/0605679].
  %%CITATION = ASTRO-PH 0605679;%%

\bibitem{Seery06a}
  D.~Seery, J.~E.~Lidsey and M.~S.~Sloth,
  %``The inflationary trispectrum,''
  arXiv:astro-ph/0610210.
  %%CITATION = ASTRO-PH 0610210;%%

\bibitem{Seery06b}
  D.~Seery and J.~E.~Lidsey,
  %``Non-gaussianity from the inflationary trispectrum,''
  arXiv:astro-ph/0611034.
  %%CITATION = ASTRO-PH 0611034;%%

\bibitem{Maldacena02}
J.~Maldacena,
%``Non-Gaussian features of primordial fluctuations
%in single field inflationary models,''
JHEP {\bf 0305}, 013 (2003).

\bibitem{Seery05a}
  D.~Seery and J.~E.~Lidsey,
  %``Primordial non-gaussianities in single field inflation,''
  JCAP {\bf 0506}, 003 (2005)
  [arXiv:astro-ph/0503692].
  %%CITATION = ASTRO-PH 0503692;%%

\bibitem{Seery05b}
  D.~Seery and J.~E.~Lidsey,
  %``Primordial non-gaussianities from multiple-field inflation,''
  JCAP {\bf 0509}, 011 (2005)
  [arXiv:astro-ph/0506056].
  %%CITATION = ASTRO-PH 0506056;%%

\bibitem{RS05}
G.~I.~Rigopoulos and E.~P.~S.~Shellard,
%``Stochastic fluctuations in multi-field inflation,''
J.\ Phys.\ Conf.\ Ser.\  {\bf 8} (2005) 145.

\bibitem{Rodriguez05}
 D.~H.~Lyth and Y.~Rodriguez,
  %``The inflationary prediction for primordial non-gaussianity,''
  Phys.\ Rev.\ Lett.\  {\bf 95}, 121302 (2005)
  [arXiv:astro-ph/0504045].
  %%CITATION = ASTRO-PH 0504045;%%

\bibitem{Alabidi:2005qi}
  L.~Alabidi and D.~H.~Lyth,
  %``Inflation models and observation,''
  JCAP {\bf 0605}, 016 (2006)
  [arXiv:astro-ph/0510441].
  %%CITATION = ASTRO-PH 0510441;%%

\bibitem{Okamoto:2002ik}
  T.~Okamoto and W.~Hu,
  %``The Angular Trispectra of CMB Temperature and Polarization,''
  Phys.\ Rev.\ D {\bf 66}, 063008 (2002)
  [arXiv:astro-ph/0206155].
  %%CITATION = ASTRO-PH 0206155;%%

\bibitem{Langlois:2005ii}
  D.~Langlois and F.~Vernizzi,
  %``Evolution of non-linear cosmological perturbations,''
  Phys.\ Rev.\ Lett.\  {\bf 95}, 091303 (2005)
  [arXiv:astro-ph/0503416].
  %%CITATION = ASTRO-PH 0503416;%%

\bibitem{Enqvist02}
K.~Enqvist and M.~S.~Sloth,
%``Adiabatic CMB perturbations in pre big
%bang string cosmology,''
Nucl.\ Phys.\ B {\bf 626}, 395 (2002).

\bibitem{Lyth02}
D.~H.~Lyth and D.~Wands,
%``Generating the curvature perturbation without an inflaton,''
Phys.\ Lett.\ B {\bf 524}, 5 (2002).

\bibitem{Moroi01}
T.~Moroi and T.~Takahashi,
%``Effects of cosmological moduli fields
%on cosmic microwave background,''
Phys.\ Lett.\ B {\bf 522}, 215 (2001)
[Erratum-ibid.\ B {\bf 539}, 303 (2002)].

\bibitem{Sasaki:2006kq}
  M.~Sasaki, J.~Valiviita and D.~Wands,
  %``Non-gaussianity of the primordial perturbation in the curvaton model,''
  Phys.\ Rev.\ D {\bf 74}, 103003 (2006)
  [arXiv:astro-ph/0607627].
  %%CITATION = ASTRO-PH 0607627;%%

\bibitem{BernardeauUzan}
  F.~Bernardeau and J.~P.~Uzan,
  %``Non-Gaussianity in multi-field inflation,''
  Phys.\ Rev.\ D {\bf 66}, 103506 (2002)
  [arXiv:hep-ph/0207295];
  %%CITATION = HEP-PH 0207295;%%
 %``Inflationary models inducing non-gaussian metric fluctuations,''
  Phys.\ Rev.\ D {\bf 67}, 121301 (2003)
  [arXiv:astro-ph/0209330].
  %%CITATION = ASTRO-PH 0209330;%%

\bibitem{Zaldarriaga}
  M.~Zaldarriaga,
  %``Non-Gaussianities in models with a varying inflaton decay rate,''
  Phys.\ Rev.\ D {\bf 69}, 043508 (2004)
  [arXiv:astro-ph/0306006].
  %%CITATION = ASTRO-PH 0306006;%%

\bibitem{Kolb}
  E.~W.~Kolb, A.~Riotto and A.~Vallinotto,
  %``Non-Gaussianity from broken symmetries,''
  Phys.\ Rev.\ D {\bf 73}, 023522 (2006)
  [arXiv:astro-ph/0511198].
  %%CITATION = ASTRO-PH 0511198;%%

\bibitem{sneutrino}
K.~Hamaguchi, H.~Murayama and T.~Yanagida,
%``Leptogenesis from sneutrino-dominated early universe,''
Phys.\ Rev.\ D {\bf 65}, 043512 (2002).

\bibitem{mssm}
K.~Enqvist and A.~Mazumdar,
%``Cosmological consequences of MSSM flat directions,''
Phys.\ Rept.\  {\bf 380}, 99 (2003).

\bibitem{MWU}
K.~A.~Malik, D.~Wands and C.~Ungarelli,
%``Large-scale curvature and entropy perturbations
%for multiple fluids,''
Phys.\ Rev.\ D {\bf 67}, 063516 (2003).

\bibitem{Linde90}
  A.~D.~Linde,
  %``Particle Physics and Inflationary Cosmology,''
  arXiv:hep-th/0503203.
  %%CITATION = HEP-TH/0503203;%%

\bibitem{Mollerach90}
S.~Mollerach,
%``Isocurvature Baryon Perturbations And Inflation,''
Phys.\ Rev.\ D {\bf 42}, 313 (1990).

\bibitem{LindeMukhanov96}
A.~D.~Linde and V.~Mukhanov,
%``Nongaussian isocurvature perturbations from inflation,''
Phys.\ Rev.\ D {\bf 56}, 535 (1997).

\bibitem{Gupta}
S.~Gupta, K.~A.~Malik and D.~Wands,
%``Curvature and isocurvature perturbations
% in a three-fluid model of  curvaton decay,''
Phys.\ Rev.\ D {\bf 69}, 063513 (2004).

\bibitem{Komatsu01}
E.~Komatsu and D.~N.~Spergel,
Phys.\ Rev.\ D {\bf 63}, 063002 (2001).

\bibitem{Lyth03}
D.~H.~Lyth, C.~Ungarelli and D.~Wands,
%``The primordial density perturbation in the curvaton scenario,''
Phys.\ Rev.\ D {\bf 67}, 023503 (2003).

\bibitem{Bartolo04cur}
N.~Bartolo, S.~Matarrese and A.~Riotto,
%``On nonGaussianity in the curvaton scenario,''
Phys.\ Rev.\ D {\bf 69}, 043503 (2004).

\bibitem{ML06}
  K.~A.~Malik and D.~H.~Lyth,
  %``A numerical study of non-gaussianity in the curvaton scenario,''
  JCAP {\bf 0609}, 008 (2006)
  [arXiv:astro-ph/0604387].
  %%CITATION = ASTRO-PH 0604387;%%

\bibitem{Cooray06}
  A.~Cooray,
  %``21-cm Background Anisotropies Can Discern Primordial Non-Gaussianity from
  %Slow-Roll Inflation,''
  arXiv:astro-ph/0610257.
  %%CITATION = ASTRO-PH 0610257;%%

\bibitem{Lewis06}
  A.~Lewis,
  %``Observational constraints and cosmological parameters,''
  arXiv:astro-ph/0603753.
  %%CITATION = ASTRO-PH 0603753;%%

\bibitem{Gordon02}
C.~Gordon and A.~Lewis,
%``Observational constraints on the curvaton model of inflation,''
Phys.\ Rev.\ D {\bf 67}, 123513 (2003).


\end{thebibliography}

%%%%%%%%%%%%%%%%%%%%%%%%%%%%%%%%%%%%%%%%%%%%%%%%%%%%%%%%%%%%%%%%%%%%%%

\printindex
\end{document}